\documentclass[%
 reprint,
 superscriptaddress,
 amsmath,amssymb,
 aps,
prl,
floatfix,
]{revtex4-1}

\usepackage{graphicx}
\usepackage{dcolumn}
\usepackage{bm}
\usepackage[normalem]{ulem}
\usepackage{hyperref}
\hypersetup{colorlinks = true, citecolor = blue, breaklinks = true}

\usepackage[usenames,dvipsnames]{xcolor}
\usepackage{textcomp}
\usepackage{multirow}
\usepackage{xfrac}

\usepackage[resetlabels]{multibib}
\newcites{SI}{References}

\usepackage{placeins}

\begin{document}

\title{Local polarization in oxygen-deficient LaMnO$_3$ induced by charge localization in the Jahn-Teller distorted structure}

\affiliation{%
Department of Chemistry and Biochemistry, University of Bern, Freiestrasse 3, CH-3012 Bern, Switzerland 
}%
\affiliation{%
National Centre for Computational Design and Discovery of Novel Materials (MARVEL), Switzerland
}%

\author{Chiara Ricca}
\affiliation{%
Department of Chemistry and Biochemistry, University of Bern, Freiestrasse 3, CH-3012 Bern, Switzerland 
}%
\affiliation{%
National Centre for Computational Design and Discovery of Novel Materials (MARVEL), Switzerland
}%

\author{Nicolas Niederhauser}
\affiliation{%
Department of Chemistry and Biochemistry, University of Bern, Freiestrasse 3, CH-3012 Bern, Switzerland 
}%

\author{Ulrich Aschauer}
\email{ulrich.aschauer@dcb.unibe.ch}
\affiliation{%
Department of Chemistry and Biochemistry, University of Bern, Freiestrasse 3, CH-3012 Bern, Switzerland 
}%
\affiliation{%
National Centre for Computational Design and Discovery of Novel Materials (MARVEL), Switzerland
}%

\date{\today}

\begin{abstract}
The functional properties of transition metal perovskite oxides are known to result from a complex interplay of magnetism, polarization, strain, and stoichiometry. Here, we show that for materials with a cooperative Jahn-Teller distortion, such as LaMnO$_3$ (LMO), the orbital order can also couple to the defect chemistry and induce novel material properties. At low temperatures, LMO exhibits a strong Jahn-Teller distortion that splits the $e_g$ orbitals of the high-spin Mn$^{3+}$ ions and leads to alternating long, short, and intermediate Mn--O bonds. Our DFT+$U$ calculations show that, as a result of this orbital order, the charge localization in LMO upon oxygen vacancy formation differs from other manganites, like SrMnO$_3$, where the two extra electrons reduce the two Mn sites adjacent to the vacancy. In LMO, relaxations around the defect depend on which type of Mn--O bond is broken, affecting the $d$-orbital energies and leading to asymmetric and hence polar excess-electron localization with respect to the vacancy. Moreover, we show that the Mn--O bond lengths, orbital order and consequently the charge localization and polarity are tunable via strain. 
\end{abstract}

\maketitle
%
The family of doped (La,Ca/Sr)MnO$_3$ perovskite manganites has attracted great interest due to its very rich phase diagram and its unusual functional properties (colossal magnetoresistance~\cite{Jin1994, Schiffer1995, millis1998lattice, Coey1999, Yamada2004}, photo-induced infrared absorption~\cite{Mertelj2000}, and efficient hole conductivity~\cite{Jiang2008}) with strong potential for new applications in the fields of electronics, spintronics, and energy conversion. The orbital order plays a significant role in determining these properties and is strongly coupled with the other electronic, structural, and spin degrees of freedom~\cite{Dagotto257, cheong2007transition}. LaMnO$_3$ (LMO) as the end-member of this family also exhibits interesting properties such as a pressure-induced insulator-to-metal transition~\cite{Loa2001} and a dielectric anomaly~\cite{Mondal2007}, that also depend on the orbital order \cite{Munoz2004,LeeBousquet2013}.

\begin{figure}[t!]
 \centering
 \includegraphics[width=0.95\columnwidth]{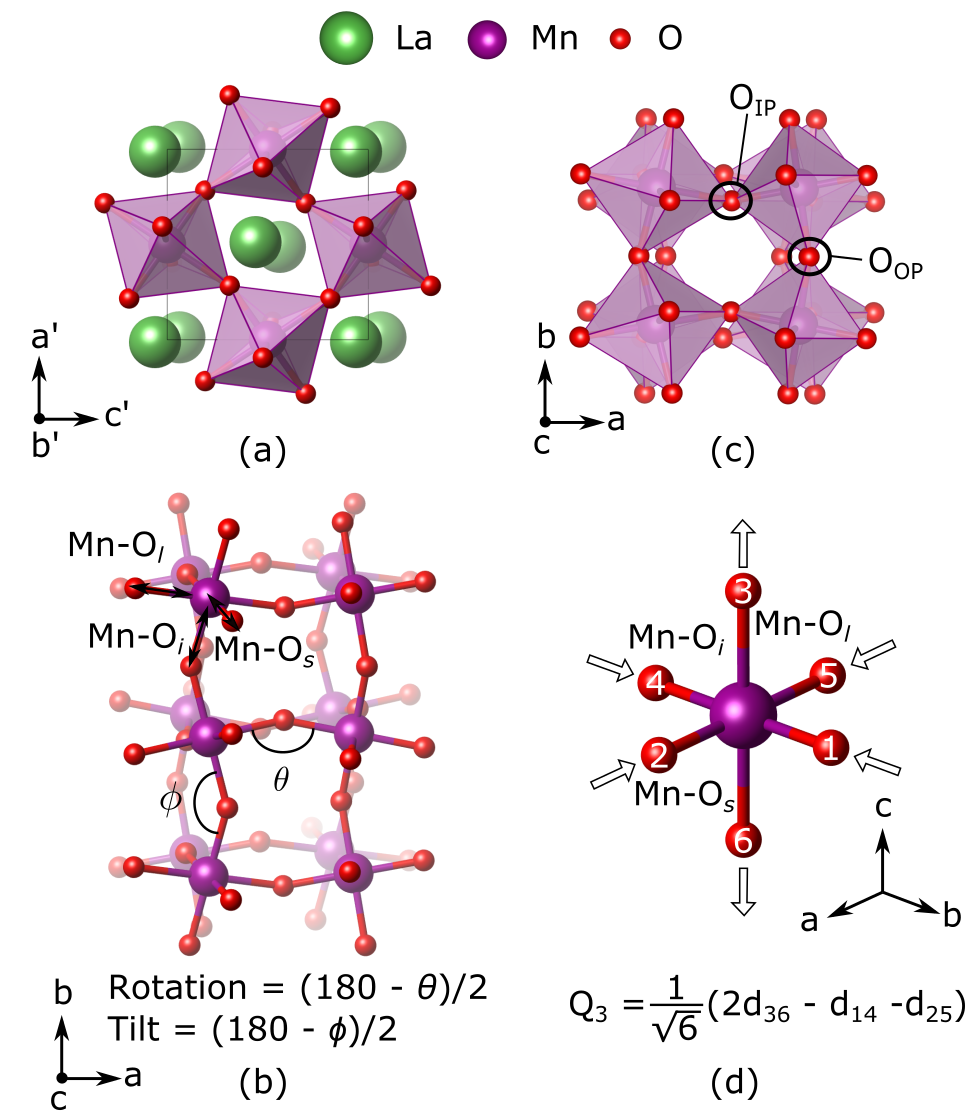}
 \caption{(a) $Pbnm$ unit cell of A-AFM LaMnO$_3$ (LMO). (b) Definition of the rotation and tilt angles and of the short (Mn--O$_s$), intermediate (Mn--O$_i$), and long (Mn--O$_l$) Mn--O bond lengths.  (c) The (2$\times$2$\times$2) LMO supercell used in this work with the two inequivalent V$_\textrm{O}$ positions: in the $ac$ plane (IP) and perpendicular to it (OP). (d) Definition of the Jahn-Teller distortion amplitude $Q_3$: $d_{36}$, $d_{14}$ and $d_{25}$ are distances between the corresponding O atoms.  Axes a', b' and c' are orthorhombic, while a, b and c are pseudo cubic.}
\label{fig:structure_bulk}
\end{figure}

Below 750~K, LMO adopts a distorted orthorhombic ($Pbnm$) perovskite structure  (see Fig.~\ref{fig:structure_bulk}a) with an A-type antiferromagnetic order (A-AFM, N\'eel temperature T$_{\textrm{N}} \approx$ 140~K~\cite{Ritter1997}), with Mn atoms coupled ferromagnetic in the $ac$ plane and antiferromagnetic along the $b$ axis~\cite{NORBY1995191, Rodriguez1998, Qiu2005}. The stabilization of this orbital-ordered insulating state is a consequence of the high-spin Mn$^{3+}$ ($t_{2g}^3e_g^1$) ions inducing a strong cooperative Jahn-Teller (JT) distortion that splits the $e_g$ orbitals ($d_{z^2}$/$d_{x^2-y^2}$ orbitals are alternately occupied within the orthorhombic $ac$-plane) and lead to alternating long and short in-plane Mn--O bonds, along with intermediate bonds along the $b$ axis (see Fig.~\ref{fig:structure_bulk}b). There are two symmetry-distinct oxygen atoms in this structure: the in-plane O ($\mathrm{O_{IP}}$) in the $ac$ plane with one short and one long Mn--O bond, and the out-of-plane O ($\mathrm{O_{OP}}$) along the $b$ direction with two intermediate Mn--O bonds (see Fig.~\ref{fig:structure_bulk}c).

Unlike most other perovskite oxides, bulk LMO can exhibit oxygen superstoichiometry (LaMnO$_{3+\delta}$), which is accommodated through cation vacancies rather than oxygen interstitials~\cite{VANROOSMALEN1994106, TOPFER1997117, Ghivelder199, LAIHO20032313}. LaMnO$_{3+\delta}$ has an extremely complex magnetic phase diagram exhibiting paramagnetic, ferromagnetic (FM), canted AFM order as well as a spin-glass state as a function of temperature and O superstoichiometry~\cite{TOPFER1997117}. Topotactic low temperature reactions~\cite{ABBATTISTA1981137, HANSTEEN2004279} were shown to lead to oxygen substoiochiometric LaMnO$_{3-\delta}$ with oxygen vacancies (V$_{\textrm{O}}$), where for $0.00 \le \delta \le 0.20$, the A-AFM order persists below T$_{\textrm{N}} \approx$ 140~K~\cite{HANSTEEN2004279}. The FM behavior often observed in LMO thin films~\cite{Gupta1995,Murugavel2003,Smadici2007,Aruta2006,garcia2010,Shah2010,Zhao2013,LIU201985} was explained by Mn$^{3+}$-Mn$^{4+}$ double exchange due to the cation deficiency generally observed in these films.
It was suggested that, below a critical thickness of 6 unit cells, LMO thin films become AFM~\cite{wang2015imaging}, but several authors also reported thicker films with bulk-like AFM behavior obtained either through growth~\cite{Kim_2010,roqueta2015strain} or high-temperature annealing~\cite{Choi_2009} under conditions favoring the formation of V$_{\textrm{O}}$. V$_{\textrm{O}}$ formation is also expected to be affected by strain via the chemical expansion~\cite{Adler2001} caused by the reducing defects, tensile strain generally favoring the formation of V$_{\textrm{O}}$~\cite{Aschauer2013, marthinsen2016, Ricca2019}.

\begin{table}
\caption{Comparison of the calculated and experimental direct and indirect band gap (E$_g^d$ and E$_g^i$ in eV) and structural properties (lattice parameters \textit{a, b, c} in \AA, Mn--O bond lengths in \AA,  octahedral rotation and tilt angle  in \textdegree, and Jahn-Teller distortion magnitude $Q_3$ in Bohr, see Fig.~\ref{fig:structure_bulk}). Experimental data were taken from Ref.~\cite{ELEMANS1971238} (structure), Ref.~\cite{Arima1993} (direct gap), and Ref.~\cite{Mahendiran1995} (indirect gap). }
\centering
\begin{tabular*}{\columnwidth}{@{\extracolsep{\fill}}lll}
\hline
\hline
        & This work & Expt. \\
\hline
$a$       & 5.496 & 5.532 \\
$b$       & 5.801 & 5.742 \\
$c$       & 7.648 & 7.668 \\
Mn--O$_s$ & 1.938 & 1.903 \\
Mn--O$_i$ & 1.986 & 1.957 \\
Mn--O$_l$ & 2.192 & 2.184 \\
Tilt    & 15.41 & 12.85 \\
Rotation     & 14.21 & 11.65 \\
$Q_3$      & 0.71  & 0.78  \\
E$_g^i$   & 0.70 & 0.24  \\
E$_g^d$   & 1.07  & 1.1   \\
\hline
\hline
\end{tabular*}
\label{tbl:bulkproperties}
\end {table}

\begin{figure}
 \centering
 \includegraphics[width=0.95\columnwidth]{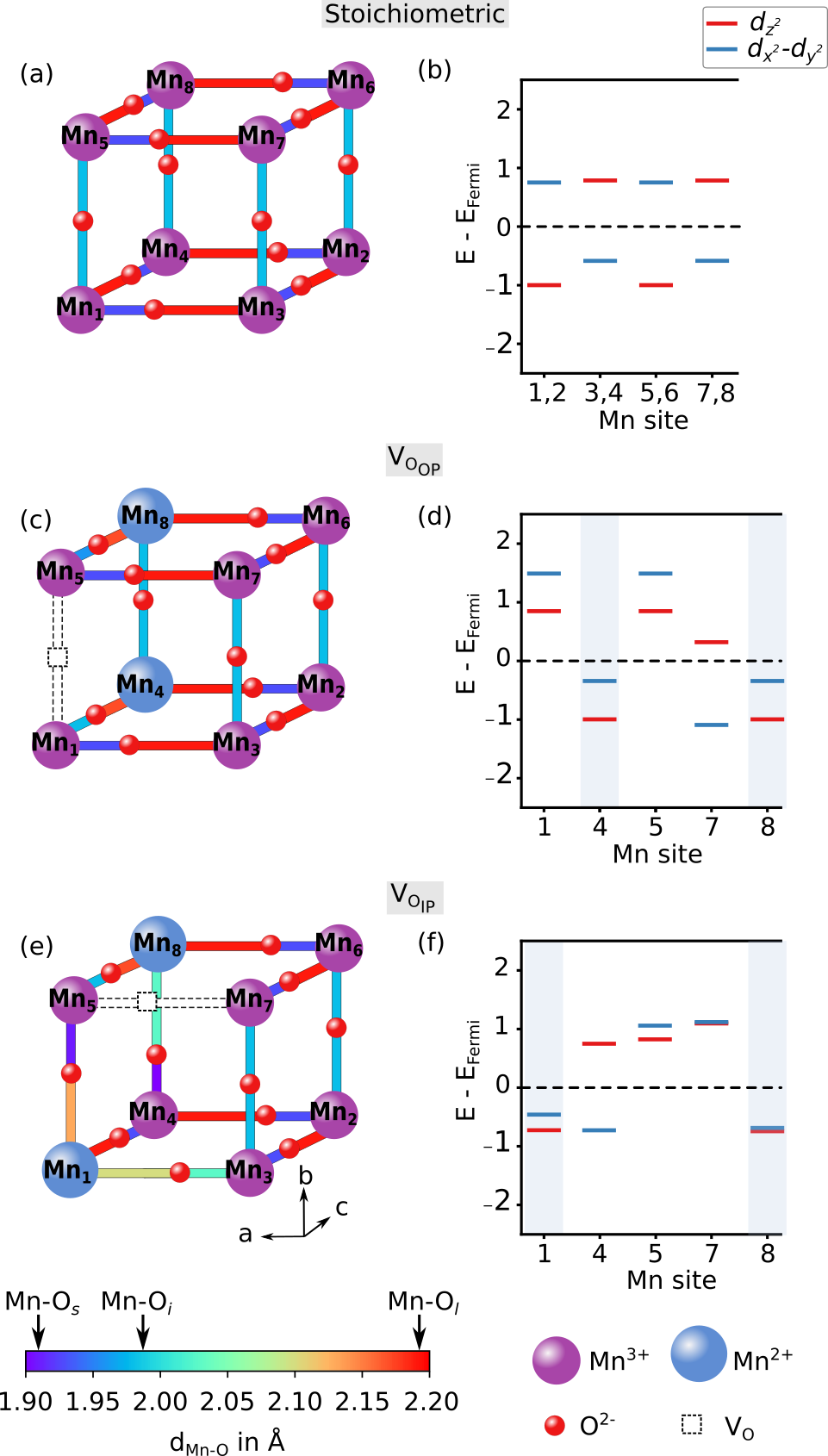}
 \caption{Schematic representation of the Mn--O bond lengths (left) and energies of the $d_{z^2}$ and $d_{x^2-y^2}$ orbitals for different Mn sites (right), determined as the centroid of the corresponding peaks in the atom- and orbital-resolved density of states. The horizontal dashed line indicates the Fermi energy. The  shaded blue area indicates sites reduced upon V$_{\textrm{O}}$ formation. Panels (a) and (b) for stoichiometric LMO, (c) and (d) for a V$_{\textrm{O}_\textrm{OP}}$ and (e) and (f) for a V$_{\textrm{O}_\textrm{IP}}$.}
\label{fig:VO_distances_orbital_changes}
\end{figure}

In this letter we investigate strained  stoichiometric and oxygen-deficient LMO by density functional theory (DFT) calculations within the {\sc{Quantum ESPRESSO}} package~\cite{giannozzi2009quantum,Giannozzi2017} (see supporting information section \ref{sec:compdetails} for details). At the PBEsol+$U$~\cite{perdew2008pbesol,Dudarev1998} level of theory with a self-consistently computed Hubbard $U$ for Mn-$3d$ states ~\cite{Timrov2018,Ricca2019}, we correctly describe the JT distorted structure, which is crucial for accurately predicted  magnetic and electronic properties~\cite{Hashimoto2010,Franchini201286,LeeBousquet2013,GAVIN201713}. As can be seen from  Table~\ref{tbl:bulkproperties}, not only are the computed lattice vectors in good agreement with experiments (relative error below 1\%), but this approach also provides a satisfactory description of the JT distortion (quantified via the magnitude of the $Q_3$ mode) and the Mn--O distances reproduced to within 2\% of experiment. We find the expected orbitally ordered insulating (see Fig.~\ref{fig:PDOS_stoich_unstrained} in the SI) state with $d_{z^2}$/$d_{x^2-y^2}$ orbitals being alternately occupied within the orthorhombic $ac$-plane, consistent with the alternating Mn-O$_s$ and Mn-O$_l$ bonds (see Figures~\ref{fig:VO_distances_orbital_changes}a and b). The direct gap of about 1 eV is in good agreement with experiments, while the indirect gap of 0.69 eV is lower than reported in previous theoretical studies~\cite{Hashimoto2010,LeeBousquet2013}, but still larger than  experiment~\cite{Mahendiran1995} (see  Table~\ref{tbl:bulkproperties}).

Upon oxygen-vacancy formation, the JT distorted structure leads to excess charge localization different from the one observed for example in CaMnO$_3$ and SrMnO$_3$ \cite{Aschauer2013, Ricca2019}, where the two excess electrons reduce Mn sites in nearest neighbor (NN) sites to the defect. We consider neutral oxygen vacancies (V$_{\textrm{O}}^{\bullet \bullet}$ in the Kr{\"o}ger-Vink notation~\cite{KROGER1956307}), but for simplicity refer to them as V$_{\textrm{O}}$ in the following. Figure~\ref{fig:VO_distances_orbital_changes}c schematically illustrates the changes in the Mn--O framework induced by the relaxations around an out-of-plane oxygen vacancy (V$_{\textrm{O}_\textrm{OP}}$). After breaking the two equivalent Mn--O$_i$ bonds along the $b$-axis, the structure relaxes by shortening the Mn--O$_l$ bonds along $c$ of the two undercoordinated NN Mn atoms (Mn$_1$ and Mn$_5$ in Fig.~\ref{fig:VO_distances_orbital_changes}c), significantly expanding the former Mn--O$_s$ bonds along that direction of the NNN Mn atoms (Mn$_4$ and Mn$_8$ in Fig.~\ref{fig:VO_distances_orbital_changes}c). As a result of these relaxations, the energy of the $d_{z^2}$ orbitals increases for Mn$_1$ and Mn$_5$ and decreases for Mn$_4$ and Mn$_8$ (see Fig.~\ref{fig:VO_distances_orbital_changes}d), localizing the two excess electrons resulting from V$_{\textrm{O}}$ formation on these orbitals and reducing the NNN Mn$_4$ and Mn$_8$ atoms (see Fig.~\ref{fig:VO_distances_orbital_changes}c). 

The case of an in-plane defect (V$_{\textrm{O}_\textrm{IP}}$), is more complex since both a Mn--O$_s$ and a Mn--O$_l$ bond are broken. As can be seen from Fig.~\ref{fig:VO_distances_orbital_changes}e, the structural relaxations upon V$_{\textrm{O}_\textrm{IP}}$ formation primarily involve one of the two NN Mn ions (Mn$_5$ and Mn$_7$ in Fig.~\ref{fig:VO_distances_orbital_changes}e where a V$_{\textrm{O}_\textrm{IP}}$ created along the $a$-axis is shown). The NN Mn atom where a Mn--O$_s$ bond was broken (Mn$_5$ in Fig.~\ref{fig:VO_distances_orbital_changes}e) shortens its remaining Mn--O bonds along the $c$ and $b$-axis, resulting in elongated Mn--O bonds for NNN Mn sites around Mn$_5$: Mn$_1$ along $b$ and Mn$_8$ along $c$, see Fig.~\ref{fig:VO_distances_orbital_changes}e. These bond-length changes are also reflected in the $d_{z^2}$ and $d_{x^2-y^2}$ orbital energies of Mn$_5$ and Mn$_7$, respectively, that are increased due to breaking the Mn$_5$--O$_\textrm{IP}$--Mn$_7$ bond, while the $d_{x^2-y^2}$ and $d_{z^2}$ orbitals of Mn$_1$ and Mn$_8$, respectively, are stabilized and filled by the excess electrons. Therefore also for the V$_{\textrm{O}_\textrm{IP}}$, the reduction happens on NNN Mn sites (see Fig.~\ref{fig:VO_distances_orbital_changes}e and f).

For both vacancies, the reduced sites show almost equivalent Mn--O bond lengths (between 2.03 and 2.17~\AA), in line with the fact that no JT distortion is expected for a Mn$^{2+}$ ion (Mn' in the Kr{\"o}ger-Vink notation~\cite{KROGER1956307} notation) in an octahedral crystal field. Furthermore, the average value of the Mn--O$_s$, Mn--O$_i$, and Mn--O$_l$ are 1.98, 2.17, and 2.20~\AA\ in good agreement with the structural parameters reported by Hasteen \textit{et al.}~\cite{HANSTEEN2004279} for LaMnO$_{2.80}$ that has a V$_{\textrm{O}}$ concentration very close to the one in our calculations. We note that this behavior was not observed in previous DFT calculations of oxygen vacancies in the high-temperature non-JT distorted G-AFM phase of LMO~\cite{Pavone2014,Olsson2016}, where the reduction of the two NN Mn atoms was reported, thus supporting that our findings are directly related to the JT distortion in the A-AFM phase.

Interestingly, when the reduction does not take place on the Mn sites adjacent to the vacancy, the localization of the excess electrons with respect to the defect can be asymmetric and hence polar. We quantify this polarization using nominal charges, using a charge of +2 on the reduced Mn identified on the base of the structural changes and  oxidation states calculated according to Ref.~\cite{Sit2011}. A small polarization of 4.06~$\mu C/cm^{2}$ in the $ac$-plane is observed for V$_{\textrm{O}_\textrm{OP}}$, in line with the reduced Mn$_4$ and Mn$_8$ (Fig.~\ref{fig:VO_distances_orbital_changes}b), being symmetrically arranged with respect to the V$_\textrm{O}$ along the $b$-axis. A larger polarization (with IP and OP components of 16.38 and 9.43~$\mu C/cm^{2}$, respectively) is computed, instead, for the V$_{\textrm{O}_\textrm{IP}}$, in line with the larger asymmetry of the charge localization observed in this case.

\begin{figure}
 \centering
 \includegraphics[width=0.95\columnwidth]{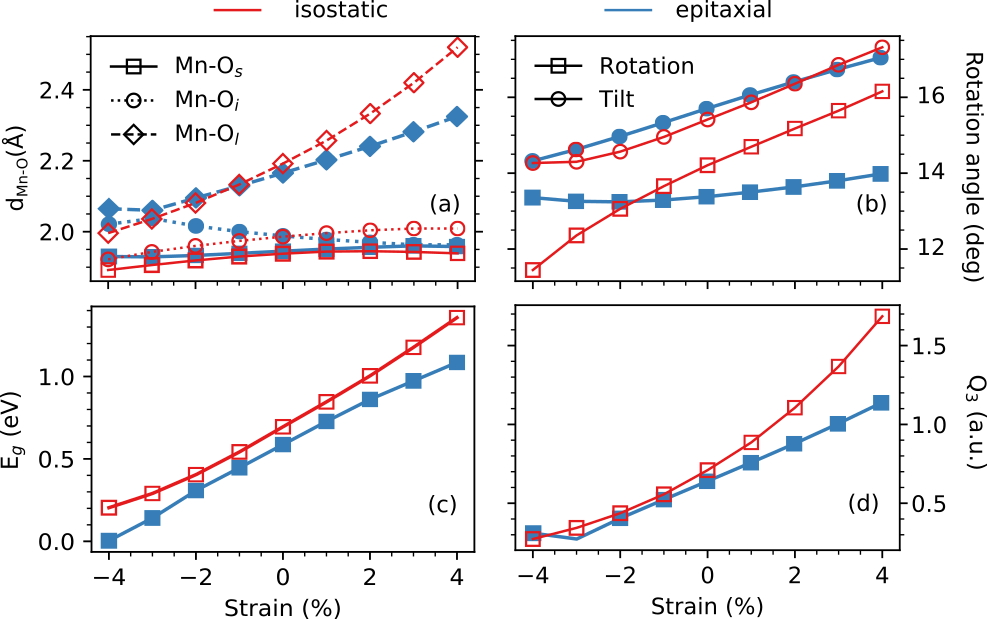}
 \caption{Evolution of structural properties with isostatic (red) and epitaxial (blue) strain in stoichiometric LMO: (a) Mn--O bond lengths, (b) octahedral rotation angles, (c) band gap, and (d) Jahn-Teller distortion magnitude $Q_3$.}
\label{fig:stoich_strain}
\end{figure}
Based on the preceding discussion, it is reasonable to expect that any external parameter affecting the Mn--O bond lengths and/or the orbital order will have a strong impact also on the charge localization and the resulting polarization. As one such external parameter, we next investigate the effect of isostatic and epitaxial strain on stoichiometric and oxygen-deficient LMO. As shown in Fig.~\ref{fig:stoich_strain}a and b isostatic and epitaxial strain in stoichiometric LMO are primarily accommodated by changes in octahedral tilt angles and bond lengths. In particular, while Mn--O$_s$ and Mn--$i$ are relatively insensitive to strain, the Mn--$l$ bonds are more strongly affected especially under tensile strain. This results in an  increase of the JT mode amplitude $Q_3$ under tensile strain (Fig.~\ref{fig:stoich_strain}c), which in turn induces an increase of the LMO band gap (Fig.~\ref{fig:stoich_strain}d). The effect of compressive strain is opposite to that of tensile strain, reducing the $Q_3$ amplitude and hence the band gap.

\begin{figure*}
 \centering
 \includegraphics[width=0.95\textwidth]{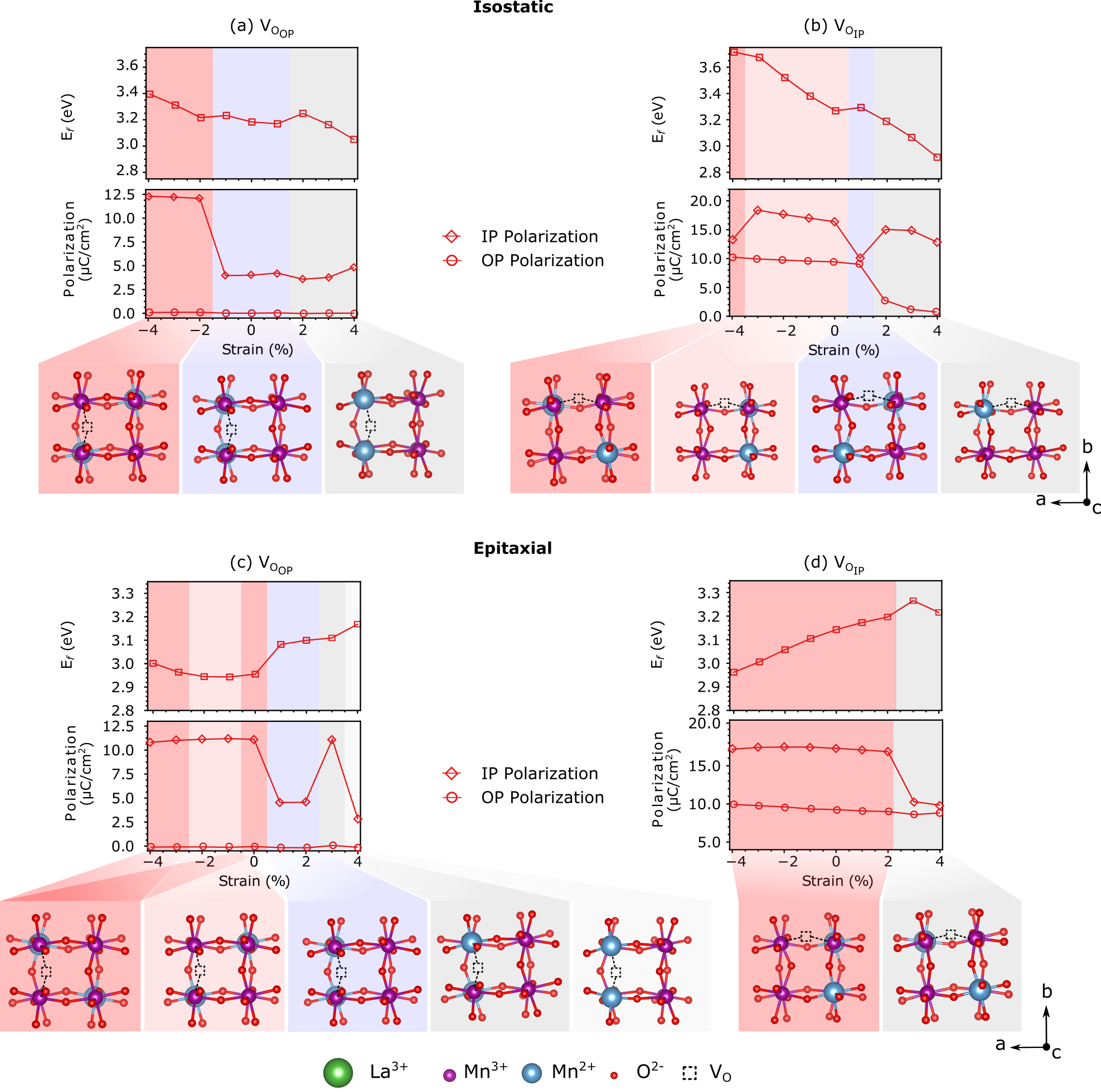}
 \caption{Oxygen-vacancy formation energy (in the oxygen-rich limit, $\Delta \mu_{\textrm{O}}= 0$, see equation~\ref{eq:formenerg} in SI) and polarization computed as a function of strain: (a) V$_{\textrm{O}_\textrm{OP}}$ under isostatic strain, (b) V$_{\textrm{O}_\textrm{IP}}$ under isostatic strain, (c) V$_{\textrm{O}_\textrm{OP}}$ under epitaxial $ac$ strain and (d) V$_{\textrm{O}_\textrm{IP}}$ under epitaxial $ac$ strain. The structures below the graphs illustrate the charge localization in the highlighted strain ranges.}
\label{fig:VO_strain}
\end{figure*}

As shown in Figure~\ref{fig:VO_strain}a, when isostatic strain is applied in presence of a V$_{\textrm{O}_\textrm{OP}}$, we observe changes in excess-charge localization as a function of the applied strain. For compressive strain larger than -2\%, one of the excess electrons localizes on a Mn further from the defect compared to the NNN sites reduced without strain. This is a consequence of the reduced Mn--O$_l$ bond lengths (see Fig.~\ref{fig:stoich_strain}). For tensile strain larger than 2\%, instead, the NN Mn sites are reduced. A similar behavior is observed for V$_{\textrm{O}_\textrm{IP}}$ (Figure~\ref{fig:VO_strain}b), but in this case, even at +4\% strain, only one of the NN Mn atoms is reduced (the one previously forming a Mn--O$_s$ bond). Despite the reduction of NNN Mn sites in some strain ranges, the formation energies of both the V$_{\textrm{O}_\textrm{OP}}$ and V$_{\textrm{O}_\textrm{IP}}$ are reduced under tensile and increased under compressive strain, with discontinuities when the excess-charge localization changes. This general trend is expected from chemical expansion arguments \cite{Aschauer2013}.

 These changes in charge localization affect also the local polarization, which depends on the relative arrangement of the two Mn'-V$_{\textrm{O}}^{\bullet \bullet}$ dipoles formed upon V$_\textrm{O}$ creation. For V$_{\textrm{O}_\textrm{OP}}$, the OP component of the polarization remains zero as the Mn' are symmetric with respect to the vacancy along the $b$ axis. The IP component is more or less constant unless when the charge localization changes, in particular under compressive isostatic strain. For V$_{\textrm{O}_\textrm{IP}}$, instead, the largest changes in polarization are observed for the OP component that almost vanishes under tensile strain due to the charge localization on one NN and one NNN Mn atom, both lying in the same atomic plane. 

Epitaxial strain, induced for example by heteroepitaxial growth on a substrate with a different lattice parameter, also affects the excess-charge localization, but the results shown in Figures~\ref{fig:VO_strain}c and d are qualitatively different compared to isostatic strain. For a V$_{\textrm{O}_\textrm{IP}}$, the charge localization changes only for tensile strain larger than 3\% (see Fig.~\ref{fig:VO_strain}d). Even at these large tensile strains, the reduction never takes place on NN but always NNN Mn sites, which is due to the non-isotropic and weaker effect on the bond lengths of epitaxial compared to isostatic strain \cite{aschauer2016interplay}. Consequently we also do not observe the chemical expansion behavior but instead the opposite trend with a reduced formation energy under compressive strain that was previously reported for ionized oxygen vacancies  \cite{aschauer2016interplay}. Moreover, at variance to isostatic strain, the IP component of the polarization is reduced for +3\% and larger strain due to the change of the relative Mn' orientation in the $ac$-plane, while the OP component stays fairly constant.

Interestingly, for V$_{\textrm{O}_\textrm{OP}}$, simply imposing the constrain of a cubic substrate with an $a$ lattice parameter resulting in the same in-plane area as LMO (0\% strain, Fig.~\ref{fig:VO_strain}c) induces a more asymmetric charge localization compared to the bulk, which is maintained also under compressive epitaxial strain. Between +1 and +2\% strain, the localization of the two excess electron changes, the reduction still taking places on NNN Mn atoms and resulting in an increase of the formation energy, similarly to V$_{\textrm{O}_\textrm{IP}}$. However, differently from the latter, for larger tensile strains one (+3\% strain) or two (+4\% strain) excess electrons localize on NN Mn sites and the formation energy remains almost constant.

In summary, we have shown that, in the case of Jahn-Teller distorted materials, orbital order is an additional parameter to take into account during defect-based design of functional material properties. As we have shown for the example of oxygen vacancies in LaMnO$_3$, unexpected excess-charge localization results from the orbital order and can lead to  the emergence of local polarization. The charge localization and hence the polarization magnitude are tunable by strain, suggesting the possibility to engineer ferroelectricity in LaMnO$_3$ when local dipoles couple for high enough defect concentrations.

\section*{\label{sec:acknow}Acknowledgments}

This research was supported by the NCCR MARVEL, funded by the Swiss National Science Foundation. Computational resources were provided by the University of Bern (on the HPC cluster UBELIX, http://www.id.unibe.ch/hpc), by the Swiss National Supercomputing Center (CSCS) under projects ID mr26 and SuperMUC at GCS@LRZ, Germany, for which we acknowledge PRACE for awarding us access.

\bibliography{references}

\begin{thebibliography}{10}%
\makeatletter
\providecommand \@ifxundefined [1]{%
 \@ifx{#1\undefined}
}%
\providecommand \@ifnum [1]{%
 \ifnum #1\expandafter \@firstoftwo
 \else \expandafter \@secondoftwo
 \fi
}%
\providecommand \@ifx [1]{%
 \ifx #1\expandafter \@firstoftwo
 \else \expandafter \@secondoftwo
 \fi
}%
\providecommand \natexlab [1]{#1}%
\providecommand \enquote  [1]{``#1''}%
\providecommand \bibnamefont  [1]{#1}%
\providecommand \bibfnamefont [1]{#1}%
\providecommand \citenamefont [1]{#1}%
\providecommand \href@noop [0]{\@secondoftwo}%
\providecommand \href [0]{\begingroup \@sanitize@url \@href}%
\providecommand \@href[1]{\@@startlink{#1}\@@href}%
\providecommand \@@href[1]{\endgroup#1\@@endlink}%
\providecommand \@sanitize@url [0]{\catcode `\\12\catcode `\$12\catcode
  `\&12\catcode `\#12\catcode `\^12\catcode `\_12\catcode `\%12\relax}%
\providecommand \@@startlink[1]{}%
\providecommand \@@endlink[0]{}%
\providecommand \url  [0]{\begingroup\@sanitize@url \@url }%
\providecommand \@url [1]{\endgroup\@href {#1}{\urlprefix }}%
\providecommand \urlprefix  [0]{URL }%
\providecommand \Eprint [0]{\href }%
\providecommand \doibase [0]{http://dx.doi.org/}%
\providecommand \selectlanguage [0]{\@gobble}%
\providecommand \bibinfo  [0]{\@secondoftwo}%
\providecommand \bibfield  [0]{\@secondoftwo}%
\providecommand \translation [1]{[#1]}%
\providecommand \BibitemOpen [0]{}%
\providecommand \bibitemStop [0]{}%
\providecommand \bibitemNoStop [0]{.\EOS\space}%
\providecommand \EOS [0]{\spacefactor3000\relax}%
\providecommand \BibitemShut  [1]{\csname bibitem#1\endcsname}%
\let\auto@bib@innerbib\@empty
\bibitem [{\citenamefont {Giannozzi}\ \emph {et~al.}(2009)\citenamefont
  {Giannozzi}, \citenamefont {Baroni}, \citenamefont {Bonini}, \citenamefont
  {Calandra}, \citenamefont {Car}, \citenamefont {Cavazzoni}, \citenamefont
  {Ceresoli}, \citenamefont {Chiarotti}, \citenamefont {Cococcioni},
  \citenamefont {Dabo}, \citenamefont {Corso}, \citenamefont {de~Gironcoli},
  \citenamefont {Fabris}, \citenamefont {Fratesi}, \citenamefont {Gebauer},
  \citenamefont {Gerstmann}, \citenamefont {Gougoussis}, \citenamefont
  {Kokalj}, \citenamefont {Lazzeri}, \citenamefont {Martin-Samos},
  \citenamefont {Marzari}, \citenamefont {Mauri}, \citenamefont {Mazzarello},
  \citenamefont {Paolini}, \citenamefont {Pasquarello}, \citenamefont
  {Paulatto}, \citenamefont {Sbraccia}, \citenamefont {Scandolo}, \citenamefont
  {Sclauzero}, \citenamefont {Seitsonen}, \citenamefont {Smogunov},
  \citenamefont {Umari},\ and\ \citenamefont
  {Wentzcovitch}}]{SI_giannozzi2009quantum}%
  \BibitemOpen
  \bibfield  {author} {\bibinfo {author} {\bibfnamefont {P.}~\bibnamefont
  {Giannozzi}}, \bibinfo {author} {\bibfnamefont {S.}~\bibnamefont {Baroni}},
  \bibinfo {author} {\bibfnamefont {N.}~\bibnamefont {Bonini}}, \bibinfo
  {author} {\bibfnamefont {M.}~\bibnamefont {Calandra}}, \bibinfo {author}
  {\bibfnamefont {R.}~\bibnamefont {Car}}, \bibinfo {author} {\bibfnamefont
  {C.}~\bibnamefont {Cavazzoni}}, \bibinfo {author} {\bibfnamefont
  {D.}~\bibnamefont {Ceresoli}}, \bibinfo {author} {\bibfnamefont {G.~L.}\
  \bibnamefont {Chiarotti}}, \bibinfo {author} {\bibfnamefont {M.}~\bibnamefont
  {Cococcioni}}, \bibinfo {author} {\bibfnamefont {I.}~\bibnamefont {Dabo}},
  \bibinfo {author} {\bibfnamefont {A.~D.}\ \bibnamefont {Corso}}, \bibinfo
  {author} {\bibfnamefont {S.}~\bibnamefont {de~Gironcoli}}, \bibinfo {author}
  {\bibfnamefont {S.}~\bibnamefont {Fabris}}, \bibinfo {author} {\bibfnamefont
  {G.}~\bibnamefont {Fratesi}}, \bibinfo {author} {\bibfnamefont
  {R.}~\bibnamefont {Gebauer}}, \bibinfo {author} {\bibfnamefont
  {U.}~\bibnamefont {Gerstmann}}, \bibinfo {author} {\bibfnamefont
  {C.}~\bibnamefont {Gougoussis}}, \bibinfo {author} {\bibfnamefont
  {A.}~\bibnamefont {Kokalj}}, \bibinfo {author} {\bibfnamefont
  {M.}~\bibnamefont {Lazzeri}}, \bibinfo {author} {\bibfnamefont
  {L.}~\bibnamefont {Martin-Samos}}, \bibinfo {author} {\bibfnamefont
  {N.}~\bibnamefont {Marzari}}, \bibinfo {author} {\bibfnamefont
  {F.}~\bibnamefont {Mauri}}, \bibinfo {author} {\bibfnamefont
  {R.}~\bibnamefont {Mazzarello}}, \bibinfo {author} {\bibfnamefont
  {S.}~\bibnamefont {Paolini}}, \bibinfo {author} {\bibfnamefont
  {A.}~\bibnamefont {Pasquarello}}, \bibinfo {author} {\bibfnamefont
  {L.}~\bibnamefont {Paulatto}}, \bibinfo {author} {\bibfnamefont
  {C.}~\bibnamefont {Sbraccia}}, \bibinfo {author} {\bibfnamefont
  {S.}~\bibnamefont {Scandolo}}, \bibinfo {author} {\bibfnamefont
  {G.}~\bibnamefont {Sclauzero}}, \bibinfo {author} {\bibfnamefont {A.~P.}\
  \bibnamefont {Seitsonen}}, \bibinfo {author} {\bibfnamefont {A.}~\bibnamefont
  {Smogunov}}, \bibinfo {author} {\bibfnamefont {P.}~\bibnamefont {Umari}}, \
  and\ \bibinfo {author} {\bibfnamefont {R.~M.}\ \bibnamefont {Wentzcovitch}},\
  }\href {\doibase 10.1088/0953-8984/21/39/395502} {\bibfield  {journal}
  {\bibinfo  {journal} {J. Phys.: Condens. Matter.}\ }\textbf {\bibinfo
  {volume} {21}},\ \bibinfo {pages} {395502} (\bibinfo {year}
  {2009})}\BibitemShut {NoStop}%
\bibitem [{\citenamefont {Giannozzi}\ \emph {et~al.}(2017)\citenamefont
  {Giannozzi}, \citenamefont {Andreussi}, \citenamefont {Brumme}, \citenamefont
  {Bunau}, \citenamefont {{Buongiorno Nardelli}}, \citenamefont {Calandra},
  \citenamefont {Car}, \citenamefont {Cavazzoni}, \citenamefont {Ceresoli},
  \citenamefont {Cococcioni}, \citenamefont {Colonna}, \citenamefont
  {Carnimeo}, \citenamefont {{Dal Corso}}, \citenamefont {{De Gironcoli}},
  \citenamefont {Delugas}, \citenamefont {Distasio}, \citenamefont {Ferretti},
  \citenamefont {Floris}, \citenamefont {Fratesi}, \citenamefont {Fugallo},
  \citenamefont {Gebauer}, \citenamefont {Gerstmann}, \citenamefont {Giustino},
  \citenamefont {Gorni}, \citenamefont {Jia}, \citenamefont {Kawamura},
  \citenamefont {Ko}, \citenamefont {Kokalj}, \citenamefont
  {K{\"{u}}c{\"{u}}kbenli}, \citenamefont {Lazzeri}, \citenamefont {Marsili},
  \citenamefont {Marzari}, \citenamefont {Mauri}, \citenamefont {Nguyen},
  \citenamefont {Nguyen}, \citenamefont {Otero-De-La-Roza}, \citenamefont
  {Paulatto}, \citenamefont {Ponc{\'{e}}}, \citenamefont {Rocca}, \citenamefont
  {Sabatini}, \citenamefont {Santra}, \citenamefont {Schlipf}, \citenamefont
  {Seitsonen}, \citenamefont {Smogunov}, \citenamefont {Timrov}, \citenamefont
  {Thonhauser}, \citenamefont {Umari}, \citenamefont {Vast}, \citenamefont
  {Wu},\ and\ \citenamefont {Baroni}}]{SI_Giannozzi2017}%
  \BibitemOpen
  \bibfield  {author} {\bibinfo {author} {\bibfnamefont {P.}~\bibnamefont
  {Giannozzi}}, \bibinfo {author} {\bibfnamefont {O.}~\bibnamefont
  {Andreussi}}, \bibinfo {author} {\bibfnamefont {T.}~\bibnamefont {Brumme}},
  \bibinfo {author} {\bibfnamefont {O.}~\bibnamefont {Bunau}}, \bibinfo
  {author} {\bibfnamefont {M.}~\bibnamefont {{Buongiorno Nardelli}}}, \bibinfo
  {author} {\bibfnamefont {M.}~\bibnamefont {Calandra}}, \bibinfo {author}
  {\bibfnamefont {R.}~\bibnamefont {Car}}, \bibinfo {author} {\bibfnamefont
  {C.}~\bibnamefont {Cavazzoni}}, \bibinfo {author} {\bibfnamefont
  {D.}~\bibnamefont {Ceresoli}}, \bibinfo {author} {\bibfnamefont
  {M.}~\bibnamefont {Cococcioni}}, \bibinfo {author} {\bibfnamefont
  {N.}~\bibnamefont {Colonna}}, \bibinfo {author} {\bibfnamefont
  {I.}~\bibnamefont {Carnimeo}}, \bibinfo {author} {\bibfnamefont
  {A.}~\bibnamefont {{Dal Corso}}}, \bibinfo {author} {\bibfnamefont
  {S.}~\bibnamefont {{De Gironcoli}}}, \bibinfo {author} {\bibfnamefont
  {P.}~\bibnamefont {Delugas}}, \bibinfo {author} {\bibfnamefont {R.~A.}\
  \bibnamefont {Distasio}}, \bibinfo {author} {\bibfnamefont {A.}~\bibnamefont
  {Ferretti}}, \bibinfo {author} {\bibfnamefont {A.}~\bibnamefont {Floris}},
  \bibinfo {author} {\bibfnamefont {G.}~\bibnamefont {Fratesi}}, \bibinfo
  {author} {\bibfnamefont {G.}~\bibnamefont {Fugallo}}, \bibinfo {author}
  {\bibfnamefont {R.}~\bibnamefont {Gebauer}}, \bibinfo {author} {\bibfnamefont
  {U.}~\bibnamefont {Gerstmann}}, \bibinfo {author} {\bibfnamefont
  {F.}~\bibnamefont {Giustino}}, \bibinfo {author} {\bibfnamefont
  {T.}~\bibnamefont {Gorni}}, \bibinfo {author} {\bibfnamefont
  {J.}~\bibnamefont {Jia}}, \bibinfo {author} {\bibfnamefont {M.}~\bibnamefont
  {Kawamura}}, \bibinfo {author} {\bibfnamefont {H.~Y.}\ \bibnamefont {Ko}},
  \bibinfo {author} {\bibfnamefont {A.}~\bibnamefont {Kokalj}}, \bibinfo
  {author} {\bibfnamefont {E.}~\bibnamefont {K{\"{u}}c{\"{u}}kbenli}}, \bibinfo
  {author} {\bibfnamefont {M.}~\bibnamefont {Lazzeri}}, \bibinfo {author}
  {\bibfnamefont {M.}~\bibnamefont {Marsili}}, \bibinfo {author} {\bibfnamefont
  {N.}~\bibnamefont {Marzari}}, \bibinfo {author} {\bibfnamefont
  {F.}~\bibnamefont {Mauri}}, \bibinfo {author} {\bibfnamefont {N.~L.}\
  \bibnamefont {Nguyen}}, \bibinfo {author} {\bibfnamefont {H.~V.}\
  \bibnamefont {Nguyen}}, \bibinfo {author} {\bibfnamefont {A.}~\bibnamefont
  {Otero-De-La-Roza}}, \bibinfo {author} {\bibfnamefont {L.}~\bibnamefont
  {Paulatto}}, \bibinfo {author} {\bibfnamefont {S.}~\bibnamefont
  {Ponc{\'{e}}}}, \bibinfo {author} {\bibfnamefont {D.}~\bibnamefont {Rocca}},
  \bibinfo {author} {\bibfnamefont {R.}~\bibnamefont {Sabatini}}, \bibinfo
  {author} {\bibfnamefont {B.}~\bibnamefont {Santra}}, \bibinfo {author}
  {\bibfnamefont {M.}~\bibnamefont {Schlipf}}, \bibinfo {author} {\bibfnamefont
  {A.~P.}\ \bibnamefont {Seitsonen}}, \bibinfo {author} {\bibfnamefont
  {A.}~\bibnamefont {Smogunov}}, \bibinfo {author} {\bibfnamefont
  {I.}~\bibnamefont {Timrov}}, \bibinfo {author} {\bibfnamefont
  {T.}~\bibnamefont {Thonhauser}}, \bibinfo {author} {\bibfnamefont
  {P.}~\bibnamefont {Umari}}, \bibinfo {author} {\bibfnamefont
  {N.}~\bibnamefont {Vast}}, \bibinfo {author} {\bibfnamefont {X.}~\bibnamefont
  {Wu}}, \ and\ \bibinfo {author} {\bibfnamefont {S.}~\bibnamefont {Baroni}},\
  }\href {\doibase 10.1088/1361-648X/aa8f79} {\bibfield  {journal} {\bibinfo
  {journal} {J. Phys.: Condens. Matter.}\ }\textbf {\bibinfo {volume} {29}},\
  \bibinfo {pages} {465901} (\bibinfo {year} {2017})}\BibitemShut {NoStop}%
\bibitem [{\citenamefont {Perdew}\ \emph {et~al.}(2008)\citenamefont {Perdew},
  \citenamefont {Ruzsinszky}, \citenamefont {Csonka}, \citenamefont {Vydrov},
  \citenamefont {Scuseria}, \citenamefont {Constantin}, \citenamefont {Zhou},\
  and\ \citenamefont {Burke}}]{SI_perdew2008pbesol}%
  \BibitemOpen
  \bibfield  {author} {\bibinfo {author} {\bibfnamefont {J.~P.}\ \bibnamefont
  {Perdew}}, \bibinfo {author} {\bibfnamefont {A.}~\bibnamefont {Ruzsinszky}},
  \bibinfo {author} {\bibfnamefont {G.~I.}\ \bibnamefont {Csonka}}, \bibinfo
  {author} {\bibfnamefont {O.~A.}\ \bibnamefont {Vydrov}}, \bibinfo {author}
  {\bibfnamefont {G.~E.}\ \bibnamefont {Scuseria}}, \bibinfo {author}
  {\bibfnamefont {L.~A.}\ \bibnamefont {Constantin}}, \bibinfo {author}
  {\bibfnamefont {X.}~\bibnamefont {Zhou}}, \ and\ \bibinfo {author}
  {\bibfnamefont {K.}~\bibnamefont {Burke}},\ }\href {\doibase
  10.1103/PhysRevLett.100.136406} {\bibfield  {journal} {\bibinfo  {journal}
  {Phys. Rev. Lett.}\ }\textbf {\bibinfo {volume} {100}},\ \bibinfo {pages}
  {136406} (\bibinfo {year} {2008})}\BibitemShut {NoStop}%
\bibitem [{\citenamefont {Vanderbilt}(1990)}]{SI_vanderbilt1990soft}%
  \BibitemOpen
  \bibfield  {author} {\bibinfo {author} {\bibfnamefont {D.}~\bibnamefont
  {Vanderbilt}},\ }\href {\doibase 10.1103/PhysRevB.41.7892} {\bibfield
  {journal} {\bibinfo  {journal} {Phys. Rev. B}\ }\textbf {\bibinfo {volume}
  {41}},\ \bibinfo {pages} {7892(R)} (\bibinfo {year} {1990})}\BibitemShut
  {NoStop}%
\bibitem [{\citenamefont {Rondinelli}\ and\ \citenamefont
  {Spaldin}(2011)}]{SI_Rondinelli:2011jk}%
  \BibitemOpen
  \bibfield  {author} {\bibinfo {author} {\bibfnamefont {J.~M.}\ \bibnamefont
  {Rondinelli}}\ and\ \bibinfo {author} {\bibfnamefont {N.~A.}\ \bibnamefont
  {Spaldin}},\ }\href {\doibase doi:10.1002/adma.201101152} {\bibfield
  {journal} {\bibinfo  {journal} {Adv. Mater.}\ }\textbf {\bibinfo {volume}
  {23}},\ \bibinfo {pages} {3363} (\bibinfo {year} {2011})}\BibitemShut
  {NoStop}%
\bibitem [{\citenamefont {Dudarev}\ \emph {et~al.}(1998)\citenamefont
  {Dudarev}, \citenamefont {Botton}, \citenamefont {Savrasov}, \citenamefont
  {Humphreys},\ and\ \citenamefont {Sutton}}]{SI_Dudarev1998}%
  \BibitemOpen
  \bibfield  {author} {\bibinfo {author} {\bibfnamefont {S.~L.}\ \bibnamefont
  {Dudarev}}, \bibinfo {author} {\bibfnamefont {G.~A.}\ \bibnamefont {Botton}},
  \bibinfo {author} {\bibfnamefont {S.~Y.}\ \bibnamefont {Savrasov}}, \bibinfo
  {author} {\bibfnamefont {C.~J.}\ \bibnamefont {Humphreys}}, \ and\ \bibinfo
  {author} {\bibfnamefont {A.~P.}\ \bibnamefont {Sutton}},\ }\href {\doibase
  10.1103/PhysRevB.57.1505} {\bibfield  {journal} {\bibinfo  {journal} {Phys.
  Rev. B}\ }\textbf {\bibinfo {volume} {57}},\ \bibinfo {pages} {1505}
  (\bibinfo {year} {1998})}\BibitemShut {NoStop}%
\bibitem [{\citenamefont {Timrov}\ \emph {et~al.}(2018)\citenamefont {Timrov},
  \citenamefont {Marzari},\ and\ \citenamefont {Cococcioni}}]{SI_Timrov2018}%
  \BibitemOpen
  \bibfield  {author} {\bibinfo {author} {\bibfnamefont {I.}~\bibnamefont
  {Timrov}}, \bibinfo {author} {\bibfnamefont {N.}~\bibnamefont {Marzari}}, \
  and\ \bibinfo {author} {\bibfnamefont {M.}~\bibnamefont {Cococcioni}},\
  }\href {\doibase 10.1103/PhysRevB.98.085127} {\bibfield  {journal} {\bibinfo
  {journal} {Phys. Rev. B}\ }\textbf {\bibinfo {volume} {98}},\ \bibinfo
  {pages} {085127} (\bibinfo {year} {2018})}\BibitemShut {NoStop}%
\bibitem [{\citenamefont {Ricca}\ \emph {et~al.}(2019)\citenamefont {Ricca},
  \citenamefont {Timrov}, \citenamefont {Cococcioni}, \citenamefont {Marzari},\
  and\ \citenamefont {Aschauer}}]{SI_Ricca2019}%
  \BibitemOpen
  \bibfield  {author} {\bibinfo {author} {\bibfnamefont {C.}~\bibnamefont
  {Ricca}}, \bibinfo {author} {\bibfnamefont {I.}~\bibnamefont {Timrov}},
  \bibinfo {author} {\bibfnamefont {M.}~\bibnamefont {Cococcioni}}, \bibinfo
  {author} {\bibfnamefont {N.}~\bibnamefont {Marzari}}, \ and\ \bibinfo
  {author} {\bibfnamefont {U.}~\bibnamefont {Aschauer}},\ }\href {\doibase
  10.1103/PhysRevB.99.094102} {\bibfield  {journal} {\bibinfo  {journal} {Phys.
  Rev. B}\ }\textbf {\bibinfo {volume} {99}},\ \bibinfo {pages} {094102}
  (\bibinfo {year} {2019})}\BibitemShut {NoStop}%
\bibitem [{\citenamefont {Freysoldt}\ \emph {et~al.}(2014)\citenamefont
  {Freysoldt}, \citenamefont {Grabowski}, \citenamefont {Hickel}, \citenamefont
  {Neugebauer}, \citenamefont {Kresse}, \citenamefont {Janotti},\ and\
  \citenamefont {Van~de Walle}}]{SI_freysoldt2014first}%
  \BibitemOpen
  \bibfield  {author} {\bibinfo {author} {\bibfnamefont {{\relax
  Ch}.}~\bibnamefont {Freysoldt}}, \bibinfo {author} {\bibfnamefont
  {B.}~\bibnamefont {Grabowski}}, \bibinfo {author} {\bibfnamefont
  {T.}~\bibnamefont {Hickel}}, \bibinfo {author} {\bibfnamefont
  {J.}~\bibnamefont {Neugebauer}}, \bibinfo {author} {\bibfnamefont
  {G.}~\bibnamefont {Kresse}}, \bibinfo {author} {\bibfnamefont
  {A.}~\bibnamefont {Janotti}}, \ and\ \bibinfo {author} {\bibfnamefont
  {C.~G.}\ \bibnamefont {Van~de Walle}},\ }\href {\doibase
  10.1103/RevModPhys.86.253} {\bibfield  {journal} {\bibinfo  {journal} {Rev.
  Mod. Phys.}\ }\textbf {\bibinfo {volume} {86}},\ \bibinfo {pages} {253}
  (\bibinfo {year} {2014})}\BibitemShut {NoStop}%
\bibitem [{\citenamefont {Sit}\ \emph {et~al.}(2011)\citenamefont {Sit},
  \citenamefont {Car}, \citenamefont {Cohen},\ and\ \citenamefont
  {Selloni}}]{SI_Sit2011}%
  \BibitemOpen
  \bibfield  {author} {\bibinfo {author} {\bibfnamefont {P.~H.-L.}\
  \bibnamefont {Sit}}, \bibinfo {author} {\bibfnamefont {R.}~\bibnamefont
  {Car}}, \bibinfo {author} {\bibfnamefont {M.~H.}\ \bibnamefont {Cohen}}, \
  and\ \bibinfo {author} {\bibfnamefont {A.}~\bibnamefont {Selloni}},\ }\href
  {\doibase 10.1021/ic2013107} {\bibfield  {journal} {\bibinfo  {journal}
  {Inorg. Chem.}\ }\textbf {\bibinfo {volume} {50}},\ \bibinfo {pages} {10259}
  (\bibinfo {year} {2011})}\BibitemShut {NoStop}%
\end{thebibliography}%


\begin{thebibliography}{55}%
\makeatletter
\providecommand \@ifxundefined [1]{%
 \@ifx{#1\undefined}
}%
\providecommand \@ifnum [1]{%
 \ifnum #1\expandafter \@firstoftwo
 \else \expandafter \@secondoftwo
 \fi
}%
\providecommand \@ifx [1]{%
 \ifx #1\expandafter \@firstoftwo
 \else \expandafter \@secondoftwo
 \fi
}%
\providecommand \natexlab [1]{#1}%
\providecommand \enquote  [1]{``#1''}%
\providecommand \bibnamefont  [1]{#1}%
\providecommand \bibfnamefont [1]{#1}%
\providecommand \citenamefont [1]{#1}%
\providecommand \href@noop [0]{\@secondoftwo}%
\providecommand \href [0]{\begingroup \@sanitize@url \@href}%
\providecommand \@href[1]{\@@startlink{#1}\@@href}%
\providecommand \@@href[1]{\endgroup#1\@@endlink}%
\providecommand \@sanitize@url [0]{\catcode `\\12\catcode `\$12\catcode
  `\&12\catcode `\#12\catcode `\^12\catcode `\_12\catcode `\%12\relax}%
\providecommand \@@startlink[1]{}%
\providecommand \@@endlink[0]{}%
\providecommand \url  [0]{\begingroup\@sanitize@url \@url }%
\providecommand \@url [1]{\endgroup\@href {#1}{\urlprefix }}%
\providecommand \urlprefix  [0]{URL }%
\providecommand \Eprint [0]{\href }%
\providecommand \doibase [0]{http://dx.doi.org/}%
\providecommand \selectlanguage [0]{\@gobble}%
\providecommand \bibinfo  [0]{\@secondoftwo}%
\providecommand \bibfield  [0]{\@secondoftwo}%
\providecommand \translation [1]{[#1]}%
\providecommand \BibitemOpen [0]{}%
\providecommand \bibitemStop [0]{}%
\providecommand \bibitemNoStop [0]{.\EOS\space}%
\providecommand \EOS [0]{\spacefactor3000\relax}%
\providecommand \BibitemShut  [1]{\csname bibitem#1\endcsname}%
\let\auto@bib@innerbib\@empty
\bibitem [{\citenamefont {Jin}\ \emph {et~al.}(1994)\citenamefont {Jin},
  \citenamefont {Tiefel}, \citenamefont {McCormack}, \citenamefont {Fastnacht},
  \citenamefont {Ramesh},\ and\ \citenamefont {Chen}}]{Jin1994}%
  \BibitemOpen
  \bibfield  {author} {\bibinfo {author} {\bibfnamefont {S.}~\bibnamefont
  {Jin}}, \bibinfo {author} {\bibfnamefont {T.~H.}\ \bibnamefont {Tiefel}},
  \bibinfo {author} {\bibfnamefont {M.}~\bibnamefont {McCormack}}, \bibinfo
  {author} {\bibfnamefont {R.~A.}\ \bibnamefont {Fastnacht}}, \bibinfo {author}
  {\bibfnamefont {R.}~\bibnamefont {Ramesh}}, \ and\ \bibinfo {author}
  {\bibfnamefont {L.~H.}\ \bibnamefont {Chen}},\ }\href {\doibase
  10.1126/science.264.5157.413} {\bibfield  {journal} {\bibinfo  {journal}
  {Science}\ }\textbf {\bibinfo {volume} {264}},\ \bibinfo {pages} {413}
  (\bibinfo {year} {1994})}\BibitemShut {NoStop}%
\bibitem [{\citenamefont {Schiffer}\ \emph {et~al.}(1995)\citenamefont
  {Schiffer}, \citenamefont {Ramirez}, \citenamefont {Bao},\ and\ \citenamefont
  {Cheong}}]{Schiffer1995}%
  \BibitemOpen
  \bibfield  {author} {\bibinfo {author} {\bibfnamefont {P.}~\bibnamefont
  {Schiffer}}, \bibinfo {author} {\bibfnamefont {A.~P.}\ \bibnamefont
  {Ramirez}}, \bibinfo {author} {\bibfnamefont {W.}~\bibnamefont {Bao}}, \ and\
  \bibinfo {author} {\bibfnamefont {S.-W.}\ \bibnamefont {Cheong}},\ }\href
  {\doibase 10.1103/PhysRevLett.75.3336} {\bibfield  {journal} {\bibinfo
  {journal} {Phys. Rev. Lett.}\ }\textbf {\bibinfo {volume} {75}},\ \bibinfo
  {pages} {3336} (\bibinfo {year} {1995})}\BibitemShut {NoStop}%
\bibitem [{\citenamefont {Millis}(1998)}]{millis1998lattice}%
  \BibitemOpen
  \bibfield  {author} {\bibinfo {author} {\bibfnamefont {A.}~\bibnamefont
  {Millis}},\ }\href {\doibase 10.1038/32348} {\bibfield  {journal} {\bibinfo
  {journal} {Nature}\ }\textbf {\bibinfo {volume} {392}},\ \bibinfo {pages}
  {147} (\bibinfo {year} {1998})}\BibitemShut {NoStop}%
\bibitem [{\citenamefont {Coey}\ \emph {et~al.}(1999)\citenamefont {Coey},
  \citenamefont {Viret},\ and\ \citenamefont {von Moln\'ar}}]{Coey1999}%
  \BibitemOpen
  \bibfield  {author} {\bibinfo {author} {\bibfnamefont {J.~M.~D.}\
  \bibnamefont {Coey}}, \bibinfo {author} {\bibfnamefont {M.}~\bibnamefont
  {Viret}}, \ and\ \bibinfo {author} {\bibfnamefont {S.}~\bibnamefont {von
  Moln\'ar}},\ }\href {\doibase 10.1080/000187399243455} {\bibfield  {journal}
  {\bibinfo  {journal} {Adv. Phys.}\ }\textbf {\bibinfo {volume} {48}},\
  \bibinfo {pages} {167} (\bibinfo {year} {1999})}\BibitemShut {NoStop}%
\bibitem [{\citenamefont {Yamada}\ \emph {et~al.}(2004)\citenamefont {Yamada},
  \citenamefont {Ogawa}, \citenamefont {Ishii}, \citenamefont {Sato},
  \citenamefont {Kawasaki}, \citenamefont {Akoh},\ and\ \citenamefont
  {Tokura}}]{Yamada2004}%
  \BibitemOpen
  \bibfield  {author} {\bibinfo {author} {\bibfnamefont {H.}~\bibnamefont
  {Yamada}}, \bibinfo {author} {\bibfnamefont {Y.}~\bibnamefont {Ogawa}},
  \bibinfo {author} {\bibfnamefont {Y.}~\bibnamefont {Ishii}}, \bibinfo
  {author} {\bibfnamefont {H.}~\bibnamefont {Sato}}, \bibinfo {author}
  {\bibfnamefont {M.}~\bibnamefont {Kawasaki}}, \bibinfo {author}
  {\bibfnamefont {H.}~\bibnamefont {Akoh}}, \ and\ \bibinfo {author}
  {\bibfnamefont {Y.}~\bibnamefont {Tokura}},\ }\href {\doibase
  10.1126/science.1098867} {\bibfield  {journal} {\bibinfo  {journal}
  {Science}\ }\textbf {\bibinfo {volume} {305}},\ \bibinfo {pages} {646}
  (\bibinfo {year} {2004})}\BibitemShut {NoStop}%
\bibitem [{\citenamefont {Mertelj}\ \emph {et~al.}(2000)\citenamefont
  {Mertelj}, \citenamefont {Ku\ifmmode \check{s}\else
  \v{s}\fi{}\ifmmode~\check{c}\else \v{c}\fi{}er}, \citenamefont {Kosec},\ and\
  \citenamefont {Mihailovic}}]{Mertelj2000}%
  \BibitemOpen
  \bibfield  {author} {\bibinfo {author} {\bibfnamefont {T.}~\bibnamefont
  {Mertelj}}, \bibinfo {author} {\bibfnamefont {D.}~\bibnamefont {Ku\ifmmode
  \check{s}\else \v{s}\fi{}\ifmmode~\check{c}\else \v{c}\fi{}er}}, \bibinfo
  {author} {\bibfnamefont {M.}~\bibnamefont {Kosec}}, \ and\ \bibinfo {author}
  {\bibfnamefont {D.}~\bibnamefont {Mihailovic}},\ }\href {\doibase
  10.1103/PhysRevB.61.15102} {\bibfield  {journal} {\bibinfo  {journal} {Phys.
  Rev. B}\ }\textbf {\bibinfo {volume} {61}},\ \bibinfo {pages} {15102}
  (\bibinfo {year} {2000})}\BibitemShut {NoStop}%
\bibitem [{\citenamefont {Jiang}(2008)}]{Jiang2008}%
  \BibitemOpen
  \bibfield  {author} {\bibinfo {author} {\bibfnamefont {S.~P.}\ \bibnamefont
  {Jiang}},\ }\href {\doibase 10.1007/s10853-008-2966-6} {\bibfield  {journal}
  {\bibinfo  {journal} {J. Mater. Sci.}\ }\textbf {\bibinfo {volume} {43}},\
  \bibinfo {pages} {6799} (\bibinfo {year} {2008})}\BibitemShut {NoStop}%
\bibitem [{\citenamefont {Dagotto}(2005)}]{Dagotto257}%
  \BibitemOpen
  \bibfield  {author} {\bibinfo {author} {\bibfnamefont {E.}~\bibnamefont
  {Dagotto}},\ }\href {\doibase 10.1126/science.1107559} {\bibfield  {journal}
  {\bibinfo  {journal} {Science}\ }\textbf {\bibinfo {volume} {309}},\ \bibinfo
  {pages} {257} (\bibinfo {year} {2005})}\BibitemShut {NoStop}%
\bibitem [{\citenamefont {Cheong}(2007)}]{cheong2007transition}%
  \BibitemOpen
  \bibfield  {author} {\bibinfo {author} {\bibfnamefont {S.-W.}\ \bibnamefont
  {Cheong}},\ }\href {\doibase doi:10.1038/nmat2069} {\bibfield  {journal}
  {\bibinfo  {journal} {Nat. Mater.}\ }\textbf {\bibinfo {volume} {6}},\
  \bibinfo {pages} {927} (\bibinfo {year} {2007})}\BibitemShut {NoStop}%
\bibitem [{\citenamefont {Loa}\ \emph {et~al.}(2001)\citenamefont {Loa},
  \citenamefont {Adler}, \citenamefont {Grzechnik}, \citenamefont {Syassen},
  \citenamefont {Schwarz}, \citenamefont {Hanfland}, \citenamefont {Rozenberg},
  \citenamefont {Gorodetsky},\ and\ \citenamefont {Pasternak}}]{Loa2001}%
  \BibitemOpen
  \bibfield  {author} {\bibinfo {author} {\bibfnamefont {I.}~\bibnamefont
  {Loa}}, \bibinfo {author} {\bibfnamefont {P.}~\bibnamefont {Adler}}, \bibinfo
  {author} {\bibfnamefont {A.}~\bibnamefont {Grzechnik}}, \bibinfo {author}
  {\bibfnamefont {K.}~\bibnamefont {Syassen}}, \bibinfo {author} {\bibfnamefont
  {U.}~\bibnamefont {Schwarz}}, \bibinfo {author} {\bibfnamefont
  {M.}~\bibnamefont {Hanfland}}, \bibinfo {author} {\bibfnamefont {G.~{\relax
  Kh}.}\ \bibnamefont {Rozenberg}}, \bibinfo {author} {\bibfnamefont
  {P.}~\bibnamefont {Gorodetsky}}, \ and\ \bibinfo {author} {\bibfnamefont
  {M.~P.}\ \bibnamefont {Pasternak}},\ }\href {\doibase
  10.1103/PhysRevLett.87.125501} {\bibfield  {journal} {\bibinfo  {journal}
  {Phys. Rev. Lett.}\ }\textbf {\bibinfo {volume} {87}},\ \bibinfo {pages}
  {125501} (\bibinfo {year} {2001})}\BibitemShut {NoStop}%
\bibitem [{\citenamefont {Mondal}\ \emph {et~al.}(2007)\citenamefont {Mondal},
  \citenamefont {Bhattacharya}, \citenamefont {Choudhury},\ and\ \citenamefont
  {Mandal}}]{Mondal2007}%
  \BibitemOpen
  \bibfield  {author} {\bibinfo {author} {\bibfnamefont {P.}~\bibnamefont
  {Mondal}}, \bibinfo {author} {\bibfnamefont {D.}~\bibnamefont
  {Bhattacharya}}, \bibinfo {author} {\bibfnamefont {P.}~\bibnamefont
  {Choudhury}}, \ and\ \bibinfo {author} {\bibfnamefont {P.}~\bibnamefont
  {Mandal}},\ }\href {\doibase 10.1103/PhysRevB.76.172403} {\bibfield
  {journal} {\bibinfo  {journal} {Phys. Rev. B}\ }\textbf {\bibinfo {volume}
  {76}},\ \bibinfo {pages} {172403} (\bibinfo {year} {2007})}\BibitemShut
  {NoStop}%
\bibitem [{\citenamefont {Mu\~noz}\ \emph {et~al.}(2004)\citenamefont
  {Mu\~noz}, \citenamefont {Harrison},\ and\ \citenamefont
  {Illas}}]{Munoz2004}%
  \BibitemOpen
  \bibfield  {author} {\bibinfo {author} {\bibfnamefont {D.}~\bibnamefont
  {Mu\~noz}}, \bibinfo {author} {\bibfnamefont {N.~M.}\ \bibnamefont
  {Harrison}}, \ and\ \bibinfo {author} {\bibfnamefont {F.}~\bibnamefont
  {Illas}},\ }\href {\doibase 10.1103/PhysRevB.69.085115} {\bibfield  {journal}
  {\bibinfo  {journal} {Phys. Rev. B}\ }\textbf {\bibinfo {volume} {69}},\
  \bibinfo {pages} {085115} (\bibinfo {year} {2004})}\BibitemShut {NoStop}%
\bibitem [{\citenamefont {Lee}\ \emph {et~al.}(2013)\citenamefont {Lee},
  \citenamefont {Delaney}, \citenamefont {Bousquet}, \citenamefont {Spaldin},\
  and\ \citenamefont {Rabe}}]{LeeBousquet2013}%
  \BibitemOpen
  \bibfield  {author} {\bibinfo {author} {\bibfnamefont {J.~H.}\ \bibnamefont
  {Lee}}, \bibinfo {author} {\bibfnamefont {K.~T.}\ \bibnamefont {Delaney}},
  \bibinfo {author} {\bibfnamefont {E.}~\bibnamefont {Bousquet}}, \bibinfo
  {author} {\bibfnamefont {N.~A.}\ \bibnamefont {Spaldin}}, \ and\ \bibinfo
  {author} {\bibfnamefont {K.~M.}\ \bibnamefont {Rabe}},\ }\href {\doibase
  10.1103/PhysRevB.88.174426} {\bibfield  {journal} {\bibinfo  {journal} {Phys.
  Rev. B}\ }\textbf {\bibinfo {volume} {88}},\ \bibinfo {pages} {174426}
  (\bibinfo {year} {2013})}\BibitemShut {NoStop}%
\bibitem [{\citenamefont {Ritter}\ \emph {et~al.}(1997)\citenamefont {Ritter},
  \citenamefont {Ibarra}, \citenamefont {De~Teresa}, \citenamefont {Algarabel},
  \citenamefont {Marquina}, \citenamefont {Blasco}, \citenamefont
  {Garc\'{\i}a}, \citenamefont {Oseroff},\ and\ \citenamefont
  {Cheong}}]{Ritter1997}%
  \BibitemOpen
  \bibfield  {author} {\bibinfo {author} {\bibfnamefont {C.}~\bibnamefont
  {Ritter}}, \bibinfo {author} {\bibfnamefont {M.~R.}\ \bibnamefont {Ibarra}},
  \bibinfo {author} {\bibfnamefont {J.~M.}\ \bibnamefont {De~Teresa}}, \bibinfo
  {author} {\bibfnamefont {P.~A.}\ \bibnamefont {Algarabel}}, \bibinfo {author}
  {\bibfnamefont {C.}~\bibnamefont {Marquina}}, \bibinfo {author}
  {\bibfnamefont {J.}~\bibnamefont {Blasco}}, \bibinfo {author} {\bibfnamefont
  {J.}~\bibnamefont {Garc\'{\i}a}}, \bibinfo {author} {\bibfnamefont
  {S.}~\bibnamefont {Oseroff}}, \ and\ \bibinfo {author} {\bibfnamefont
  {S.-W.}\ \bibnamefont {Cheong}},\ }\href {\doibase 10.1103/PhysRevB.56.8902}
  {\bibfield  {journal} {\bibinfo  {journal} {Phys. Rev. B}\ }\textbf {\bibinfo
  {volume} {56}},\ \bibinfo {pages} {8902} (\bibinfo {year}
  {1997})}\BibitemShut {NoStop}%
\bibitem [{\citenamefont {Norby}\ \emph {et~al.}(1995)\citenamefont {Norby},
  \citenamefont {Krogh~Andersen}, \citenamefont {Krogh~Andersen},\ and\
  \citenamefont {Andersen}}]{NORBY1995191}%
  \BibitemOpen
  \bibfield  {author} {\bibinfo {author} {\bibfnamefont {P.}~\bibnamefont
  {Norby}}, \bibinfo {author} {\bibfnamefont {I.~G.}\ \bibnamefont
  {Krogh~Andersen}}, \bibinfo {author} {\bibfnamefont {E.}~\bibnamefont
  {Krogh~Andersen}}, \ and\ \bibinfo {author} {\bibfnamefont {N.~H.}\
  \bibnamefont {Andersen}},\ }\href {\doibase 10.1016/0022-4596(95)80028-N}
  {\bibfield  {journal} {\bibinfo  {journal} {J. Solid State Chem.}\ }\textbf
  {\bibinfo {volume} {119}},\ \bibinfo {pages} {191} (\bibinfo {year}
  {1995})}\BibitemShut {NoStop}%
\bibitem [{\citenamefont {Rodr\'{\i}guez-Carvajal}\ \emph
  {et~al.}(1998)\citenamefont {Rodr\'{\i}guez-Carvajal}, \citenamefont
  {Hennion}, \citenamefont {Moussa}, \citenamefont {Moudden}, \citenamefont
  {Pinsard},\ and\ \citenamefont {Revcolevschi}}]{Rodriguez1998}%
  \BibitemOpen
  \bibfield  {author} {\bibinfo {author} {\bibfnamefont {J.}~\bibnamefont
  {Rodr\'{\i}guez-Carvajal}}, \bibinfo {author} {\bibfnamefont
  {M.}~\bibnamefont {Hennion}}, \bibinfo {author} {\bibfnamefont
  {F.}~\bibnamefont {Moussa}}, \bibinfo {author} {\bibfnamefont {A.~H.}\
  \bibnamefont {Moudden}}, \bibinfo {author} {\bibfnamefont {L.}~\bibnamefont
  {Pinsard}}, \ and\ \bibinfo {author} {\bibfnamefont {A.}~\bibnamefont
  {Revcolevschi}},\ }\href {\doibase 10.1103/PhysRevB.57.R3189} {\bibfield
  {journal} {\bibinfo  {journal} {Phys. Rev. B}\ }\textbf {\bibinfo {volume}
  {57}},\ \bibinfo {pages} {3189(R)} (\bibinfo {year} {1998})}\BibitemShut
  {NoStop}%
\bibitem [{\citenamefont {Qiu}\ \emph {et~al.}(2005)\citenamefont {Qiu},
  \citenamefont {Proffen}, \citenamefont {Mitchell},\ and\ \citenamefont
  {Billinge}}]{Qiu2005}%
  \BibitemOpen
  \bibfield  {author} {\bibinfo {author} {\bibfnamefont {X.}~\bibnamefont
  {Qiu}}, \bibinfo {author} {\bibfnamefont {{\relax Th}.}~\bibnamefont
  {Proffen}}, \bibinfo {author} {\bibfnamefont {J.~F.}\ \bibnamefont
  {Mitchell}}, \ and\ \bibinfo {author} {\bibfnamefont {S.~J.~L.}\ \bibnamefont
  {Billinge}},\ }\href {\doibase 10.1103/PhysRevLett.94.177203} {\bibfield
  {journal} {\bibinfo  {journal} {Phys. Rev. Lett.}\ }\textbf {\bibinfo
  {volume} {94}},\ \bibinfo {pages} {177203} (\bibinfo {year}
  {2005})}\BibitemShut {NoStop}%
\bibitem [{\citenamefont {Van~Roosmalen}\ and\ \citenamefont
  {Cordfunke}(1994)}]{VANROOSMALEN1994106}%
  \BibitemOpen
  \bibfield  {author} {\bibinfo {author} {\bibfnamefont {J.~A.~M.}\
  \bibnamefont {Van~Roosmalen}}\ and\ \bibinfo {author} {\bibfnamefont
  {E.~H.~P.}\ \bibnamefont {Cordfunke}},\ }\href {\doibase
  10.1006/jssc.1994.1142} {\bibfield  {journal} {\bibinfo  {journal} {J. Solid
  State Chem.}\ }\textbf {\bibinfo {volume} {110}},\ \bibinfo {pages} {106 }
  (\bibinfo {year} {1994})}\BibitemShut {NoStop}%
\bibitem [{\citenamefont {T\"opfer}\ and\ \citenamefont
  {Goodenough}(1997)}]{TOPFER1997117}%
  \BibitemOpen
  \bibfield  {author} {\bibinfo {author} {\bibfnamefont {J.}~\bibnamefont
  {T\"opfer}}\ and\ \bibinfo {author} {\bibfnamefont {J.~B.}\ \bibnamefont
  {Goodenough}},\ }\href {\doibase 10.1006/jssc.1997.7287} {\bibfield
  {journal} {\bibinfo  {journal} {J. Solid State Chem.}\ }\textbf {\bibinfo
  {volume} {130}},\ \bibinfo {pages} {117 } (\bibinfo {year}
  {1997})}\BibitemShut {NoStop}%
\bibitem [{\citenamefont {Ghivelder}\ \emph {et~al.}(1999)\citenamefont
  {Ghivelder}, \citenamefont {Abrego~Castillo}, \citenamefont {Gusm\~ao},
  \citenamefont {Alonso},\ and\ \citenamefont {Cohen}}]{Ghivelder199}%
  \BibitemOpen
  \bibfield  {author} {\bibinfo {author} {\bibfnamefont {L.}~\bibnamefont
  {Ghivelder}}, \bibinfo {author} {\bibfnamefont {I.}~\bibnamefont
  {Abrego~Castillo}}, \bibinfo {author} {\bibfnamefont {M.~A.}\ \bibnamefont
  {Gusm\~ao}}, \bibinfo {author} {\bibfnamefont {J.~A.}\ \bibnamefont
  {Alonso}}, \ and\ \bibinfo {author} {\bibfnamefont {L.~F.}\ \bibnamefont
  {Cohen}},\ }\href {\doibase 10.1103/PhysRevB.60.12184} {\bibfield  {journal}
  {\bibinfo  {journal} {Phys. Rev. B}\ }\textbf {\bibinfo {volume} {60}},\
  \bibinfo {pages} {12184} (\bibinfo {year} {1999})}\BibitemShut {NoStop}%
\bibitem [{\citenamefont {Laiho}\ \emph {et~al.}(2003)\citenamefont {Laiho},
  \citenamefont {Lisunov}, \citenamefont {L\"ahderanta}, \citenamefont
  {Petrenko}, \citenamefont {Salminen}, \citenamefont {Stamov}, \citenamefont
  {Stepanov},\ and\ \citenamefont {Zakhvalinskii}}]{LAIHO20032313}%
  \BibitemOpen
  \bibfield  {author} {\bibinfo {author} {\bibfnamefont {R.}~\bibnamefont
  {Laiho}}, \bibinfo {author} {\bibfnamefont {K.~G.}\ \bibnamefont {Lisunov}},
  \bibinfo {author} {\bibfnamefont {E.}~\bibnamefont {L\"ahderanta}}, \bibinfo
  {author} {\bibfnamefont {P.~A.}\ \bibnamefont {Petrenko}}, \bibinfo {author}
  {\bibfnamefont {J.}~\bibnamefont {Salminen}}, \bibinfo {author}
  {\bibfnamefont {V.~N.}\ \bibnamefont {Stamov}}, \bibinfo {author}
  {\bibfnamefont {{\relax Yu}.~P.}\ \bibnamefont {Stepanov}}, \ and\ \bibinfo
  {author} {\bibfnamefont {V.~S.}\ \bibnamefont {Zakhvalinskii}},\ }\href
  {\doibase 10.1016/S0022-3697(03)00266-X} {\bibfield  {journal} {\bibinfo
  {journal} {J. Phys. Chem. Solids}\ }\textbf {\bibinfo {volume} {64}},\
  \bibinfo {pages} {2313 } (\bibinfo {year} {2003})}\BibitemShut {NoStop}%
\bibitem [{\citenamefont {Abbattista}\ and\ \citenamefont
  {Lucco~Borlera}(1981)}]{ABBATTISTA1981137}%
  \BibitemOpen
  \bibfield  {author} {\bibinfo {author} {\bibfnamefont {F.}~\bibnamefont
  {Abbattista}}\ and\ \bibinfo {author} {\bibfnamefont {M.}~\bibnamefont
  {Lucco~Borlera}},\ }\href {\doibase 10.1016/0272-8842(81)90010-9} {\bibfield
  {journal} {\bibinfo  {journal} {Ceram. Int.}\ }\textbf {\bibinfo {volume}
  {7}},\ \bibinfo {pages} {137} (\bibinfo {year} {1981})}\BibitemShut {NoStop}%
\bibitem [{\citenamefont {Hansteen}\ \emph {et~al.}(2004)\citenamefont
  {Hansteen}, \citenamefont {Bréard}, \citenamefont {Fjellvåg},\ and\
  \citenamefont {Hauback}}]{HANSTEEN2004279}%
  \BibitemOpen
  \bibfield  {author} {\bibinfo {author} {\bibfnamefont {O.~H.}\ \bibnamefont
  {Hansteen}}, \bibinfo {author} {\bibfnamefont {Y.}~\bibnamefont {Bréard}},
  \bibinfo {author} {\bibfnamefont {H.}~\bibnamefont {Fjellvåg}}, \ and\
  \bibinfo {author} {\bibfnamefont {B.~C.}\ \bibnamefont {Hauback}},\ }\href
  {\doibase 10.1016/j.solidstatesciences.2004.01.002} {\bibfield  {journal}
  {\bibinfo  {journal} {Solid State Sci.}\ }\textbf {\bibinfo {volume} {6}},\
  \bibinfo {pages} {279} (\bibinfo {year} {2004})}\BibitemShut {NoStop}%
\bibitem [{\citenamefont {Gupta}\ \emph {et~al.}(1995)\citenamefont {Gupta},
  \citenamefont {McGuire}, \citenamefont {Duncombe}, \citenamefont {Rupp},
  \citenamefont {Sun}, \citenamefont {Gallagher},\ and\ \citenamefont
  {Xiao}}]{Gupta1995}%
  \BibitemOpen
  \bibfield  {author} {\bibinfo {author} {\bibfnamefont {A.}~\bibnamefont
  {Gupta}}, \bibinfo {author} {\bibfnamefont {T.~R.}\ \bibnamefont {McGuire}},
  \bibinfo {author} {\bibfnamefont {P.~R.}\ \bibnamefont {Duncombe}}, \bibinfo
  {author} {\bibfnamefont {M.}~\bibnamefont {Rupp}}, \bibinfo {author}
  {\bibfnamefont {J.~Z.}\ \bibnamefont {Sun}}, \bibinfo {author} {\bibfnamefont
  {W.~J.}\ \bibnamefont {Gallagher}}, \ and\ \bibinfo {author} {\bibfnamefont
  {G.}~\bibnamefont {Xiao}},\ }\href {\doibase 10.1063/1.115258} {\bibfield
  {journal} {\bibinfo  {journal} {Appl. Phys. Lett.}\ }\textbf {\bibinfo
  {volume} {67}},\ \bibinfo {pages} {3494} (\bibinfo {year}
  {1995})}\BibitemShut {NoStop}%
\bibitem [{\citenamefont {Murugavel}\ \emph {et~al.}(2003)\citenamefont
  {Murugavel}, \citenamefont {Lee}, \citenamefont {Yoon}, \citenamefont {Noh},
  \citenamefont {Chung}, \citenamefont {Heu},\ and\ \citenamefont
  {Yoon}}]{Murugavel2003}%
  \BibitemOpen
  \bibfield  {author} {\bibinfo {author} {\bibfnamefont {P.}~\bibnamefont
  {Murugavel}}, \bibinfo {author} {\bibfnamefont {J.~H.}\ \bibnamefont {Lee}},
  \bibinfo {author} {\bibfnamefont {J.-G.}\ \bibnamefont {Yoon}}, \bibinfo
  {author} {\bibfnamefont {T.~W.}\ \bibnamefont {Noh}}, \bibinfo {author}
  {\bibfnamefont {J.-S.}\ \bibnamefont {Chung}}, \bibinfo {author}
  {\bibfnamefont {M.}~\bibnamefont {Heu}}, \ and\ \bibinfo {author}
  {\bibfnamefont {S.}~\bibnamefont {Yoon}},\ }\href {\doibase
  10.1063/1.1563740} {\bibfield  {journal} {\bibinfo  {journal} {Appl. Phys.
  Lett.}\ }\textbf {\bibinfo {volume} {82}},\ \bibinfo {pages} {1908} (\bibinfo
  {year} {2003})}\BibitemShut {NoStop}%
\bibitem [{\citenamefont {Smadici}\ \emph {et~al.}(2007)\citenamefont
  {Smadici}, \citenamefont {Abbamonte}, \citenamefont {Bhattacharya},
  \citenamefont {Zhai}, \citenamefont {Jiang}, \citenamefont {Rusydi},
  \citenamefont {Eckstein}, \citenamefont {Bader},\ and\ \citenamefont
  {Zuo}}]{Smadici2007}%
  \BibitemOpen
  \bibfield  {author} {\bibinfo {author} {\bibfnamefont {S.}~\bibnamefont
  {Smadici}}, \bibinfo {author} {\bibfnamefont {P.}~\bibnamefont {Abbamonte}},
  \bibinfo {author} {\bibfnamefont {A.}~\bibnamefont {Bhattacharya}}, \bibinfo
  {author} {\bibfnamefont {X.}~\bibnamefont {Zhai}}, \bibinfo {author}
  {\bibfnamefont {B.}~\bibnamefont {Jiang}}, \bibinfo {author} {\bibfnamefont
  {A.}~\bibnamefont {Rusydi}}, \bibinfo {author} {\bibfnamefont {J.~N.}\
  \bibnamefont {Eckstein}}, \bibinfo {author} {\bibfnamefont {S.~D.}\
  \bibnamefont {Bader}}, \ and\ \bibinfo {author} {\bibfnamefont {J.-M.}\
  \bibnamefont {Zuo}},\ }\href {\doibase 10.1103/PhysRevLett.99.196404}
  {\bibfield  {journal} {\bibinfo  {journal} {Phys. Rev. Lett.}\ }\textbf
  {\bibinfo {volume} {99}},\ \bibinfo {pages} {196404} (\bibinfo {year}
  {2007})}\BibitemShut {NoStop}%
\bibitem [{\citenamefont {Aruta}\ \emph {et~al.}(2006)\citenamefont {Aruta},
  \citenamefont {Angeloni}, \citenamefont {Balestrino}, \citenamefont {Boggio},
  \citenamefont {Medaglia}, \citenamefont {Tebano}, \citenamefont {Davidson},
  \citenamefont {Baldini}, \citenamefont {Di~Castro}, \citenamefont
  {Postorino}, \citenamefont {Dore}, \citenamefont {Sidorenko}, \citenamefont
  {Allodi},\ and\ \citenamefont {De~Renzi}}]{Aruta2006}%
  \BibitemOpen
  \bibfield  {author} {\bibinfo {author} {\bibfnamefont {C.}~\bibnamefont
  {Aruta}}, \bibinfo {author} {\bibfnamefont {M.}~\bibnamefont {Angeloni}},
  \bibinfo {author} {\bibfnamefont {G.}~\bibnamefont {Balestrino}}, \bibinfo
  {author} {\bibfnamefont {N.~G.}\ \bibnamefont {Boggio}}, \bibinfo {author}
  {\bibfnamefont {P.~G.}\ \bibnamefont {Medaglia}}, \bibinfo {author}
  {\bibfnamefont {A.}~\bibnamefont {Tebano}}, \bibinfo {author} {\bibfnamefont
  {B.}~\bibnamefont {Davidson}}, \bibinfo {author} {\bibfnamefont
  {M.}~\bibnamefont {Baldini}}, \bibinfo {author} {\bibfnamefont
  {D.}~\bibnamefont {Di~Castro}}, \bibinfo {author} {\bibfnamefont
  {P.}~\bibnamefont {Postorino}}, \bibinfo {author} {\bibfnamefont
  {P.}~\bibnamefont {Dore}}, \bibinfo {author} {\bibfnamefont {A.}~\bibnamefont
  {Sidorenko}}, \bibinfo {author} {\bibfnamefont {G.}~\bibnamefont {Allodi}}, \
  and\ \bibinfo {author} {\bibfnamefont {R.}~\bibnamefont {De~Renzi}},\ }\href
  {\doibase 10.1063/1.2217983} {\bibfield  {journal} {\bibinfo  {journal} {J.
  Appl. Phys.}\ }\textbf {\bibinfo {volume} {100}},\ \bibinfo {pages} {023910}
  (\bibinfo {year} {2006})}\BibitemShut {NoStop}%
\bibitem [{\citenamefont {Garcia-Barriocanal}\ \emph
  {et~al.}(2010)\citenamefont {Garcia-Barriocanal}, \citenamefont {Bruno},
  \citenamefont {Rivera-Calzada}, \citenamefont {Sefrioui}, \citenamefont
  {Nemes}, \citenamefont {Garcia-Hernández}, \citenamefont {Rubio-Zuazo},
  \citenamefont {Castro}, \citenamefont {Varela}, \citenamefont {Pennycook},
  \citenamefont {Leon},\ and\ \citenamefont {Santamaria}}]{garcia2010}%
  \BibitemOpen
  \bibfield  {author} {\bibinfo {author} {\bibfnamefont {J.}~\bibnamefont
  {Garcia-Barriocanal}}, \bibinfo {author} {\bibfnamefont {F.~Y.}\ \bibnamefont
  {Bruno}}, \bibinfo {author} {\bibfnamefont {A.}~\bibnamefont
  {Rivera-Calzada}}, \bibinfo {author} {\bibfnamefont {Z.}~\bibnamefont
  {Sefrioui}}, \bibinfo {author} {\bibfnamefont {N.~M.}\ \bibnamefont {Nemes}},
  \bibinfo {author} {\bibfnamefont {M.}~\bibnamefont {Garcia-Hernández}},
  \bibinfo {author} {\bibfnamefont {J.}~\bibnamefont {Rubio-Zuazo}}, \bibinfo
  {author} {\bibfnamefont {G.~R.}\ \bibnamefont {Castro}}, \bibinfo {author}
  {\bibfnamefont {M.}~\bibnamefont {Varela}}, \bibinfo {author} {\bibfnamefont
  {S.~J.}\ \bibnamefont {Pennycook}}, \bibinfo {author} {\bibfnamefont
  {C.}~\bibnamefont {Leon}}, \ and\ \bibinfo {author} {\bibfnamefont
  {J.}~\bibnamefont {Santamaria}},\ }\href {\doibase 10.1002/adma.200902263}
  {\bibfield  {journal} {\bibinfo  {journal} {Adv. Mat.}\ }\textbf {\bibinfo
  {volume} {22}},\ \bibinfo {pages} {627} (\bibinfo {year} {2010})}\BibitemShut
  {NoStop}%
\bibitem [{\citenamefont {Shah}\ \emph {et~al.}(2010)\citenamefont {Shah},
  \citenamefont {Ramasse}, \citenamefont {Zhai}, \citenamefont {Wen},
  \citenamefont {May}, \citenamefont {Petrov}, \citenamefont {Bhattacharya},
  \citenamefont {Abbamonte}, \citenamefont {Eckstein},\ and\ \citenamefont
  {Zuo}}]{Shah2010}%
  \BibitemOpen
  \bibfield  {author} {\bibinfo {author} {\bibfnamefont {A.~B.}\ \bibnamefont
  {Shah}}, \bibinfo {author} {\bibfnamefont {Q.~M.}\ \bibnamefont {Ramasse}},
  \bibinfo {author} {\bibfnamefont {X.}~\bibnamefont {Zhai}}, \bibinfo {author}
  {\bibfnamefont {J.~G.}\ \bibnamefont {Wen}}, \bibinfo {author} {\bibfnamefont
  {S.~J.}\ \bibnamefont {May}}, \bibinfo {author} {\bibfnamefont
  {I.}~\bibnamefont {Petrov}}, \bibinfo {author} {\bibfnamefont
  {A.}~\bibnamefont {Bhattacharya}}, \bibinfo {author} {\bibfnamefont
  {P.}~\bibnamefont {Abbamonte}}, \bibinfo {author} {\bibfnamefont {J.~N.}\
  \bibnamefont {Eckstein}}, \ and\ \bibinfo {author} {\bibfnamefont {J.-M.}\
  \bibnamefont {Zuo}},\ }\href {\doibase 10.1002/adma.200904198} {\bibfield
  {journal} {\bibinfo  {journal} {Adv. Mat.}\ }\textbf {\bibinfo {volume}
  {22}},\ \bibinfo {pages} {1156} (\bibinfo {year} {2010})}\BibitemShut
  {NoStop}%
\bibitem [{\citenamefont {Zhao}\ \emph {et~al.}(2013)\citenamefont {Zhao},
  \citenamefont {Jin}, \citenamefont {Xu}, \citenamefont {Guo}, \citenamefont
  {Wang}, \citenamefont {Ge}, \citenamefont {Lu},\ and\ \citenamefont
  {Yang}}]{Zhao2013}%
  \BibitemOpen
  \bibfield  {author} {\bibinfo {author} {\bibfnamefont {R.}~\bibnamefont
  {Zhao}}, \bibinfo {author} {\bibfnamefont {K.}~\bibnamefont {Jin}}, \bibinfo
  {author} {\bibfnamefont {Z.}~\bibnamefont {Xu}}, \bibinfo {author}
  {\bibfnamefont {H.}~\bibnamefont {Guo}}, \bibinfo {author} {\bibfnamefont
  {L.}~\bibnamefont {Wang}}, \bibinfo {author} {\bibfnamefont {C.}~\bibnamefont
  {Ge}}, \bibinfo {author} {\bibfnamefont {H.}~\bibnamefont {Lu}}, \ and\
  \bibinfo {author} {\bibfnamefont {G.}~\bibnamefont {Yang}},\ }\href {\doibase
  10.1063/1.4798550} {\bibfield  {journal} {\bibinfo  {journal} {Appl. Phys.
  Lett.}\ }\textbf {\bibinfo {volume} {102}},\ \bibinfo {pages} {122402}
  (\bibinfo {year} {2013})}\BibitemShut {NoStop}%
\bibitem [{\citenamefont {Liu}\ \emph {et~al.}(2019)\citenamefont {Liu},
  \citenamefont {Wong}, \citenamefont {Lam}, \citenamefont {Mak},\ and\
  \citenamefont {Leung}}]{LIU201985}%
  \BibitemOpen
  \bibfield  {author} {\bibinfo {author} {\bibfnamefont {Y.~K.}\ \bibnamefont
  {Liu}}, \bibinfo {author} {\bibfnamefont {H.~F.}\ \bibnamefont {Wong}},
  \bibinfo {author} {\bibfnamefont {K.~K.}\ \bibnamefont {Lam}}, \bibinfo
  {author} {\bibfnamefont {C.~L.}\ \bibnamefont {Mak}}, \ and\ \bibinfo
  {author} {\bibfnamefont {C.~W.}\ \bibnamefont {Leung}},\ }\href {\doibase
  10.1016/j.jmmm.2019.03.001} {\bibfield  {journal} {\bibinfo  {journal} {J.
  Magn. Magn. Mater.}\ }\textbf {\bibinfo {volume} {481}},\ \bibinfo {pages}
  {85} (\bibinfo {year} {2019})}\BibitemShut {NoStop}%
\bibitem [{\citenamefont {Wang}\ \emph {et~al.}(2015)\citenamefont {Wang},
  \citenamefont {Li}, \citenamefont {L{\"u}}, \citenamefont {Paudel},
  \citenamefont {Leusink}, \citenamefont {Hoek}, \citenamefont {Poccia},
  \citenamefont {Vailionis}, \citenamefont {Venkatesan}, \citenamefont {Coey},
  \citenamefont {Tsymbal}, \citenamefont {Ariando},\ and\ \citenamefont
  {Hilgenkamp}}]{wang2015imaging}%
  \BibitemOpen
  \bibfield  {author} {\bibinfo {author} {\bibfnamefont {X.~R.}\ \bibnamefont
  {Wang}}, \bibinfo {author} {\bibfnamefont {C.~J.}\ \bibnamefont {Li}},
  \bibinfo {author} {\bibfnamefont {W.~M.}\ \bibnamefont {L{\"u}}}, \bibinfo
  {author} {\bibfnamefont {T.~R.}\ \bibnamefont {Paudel}}, \bibinfo {author}
  {\bibfnamefont {D.~P.}\ \bibnamefont {Leusink}}, \bibinfo {author}
  {\bibfnamefont {M.}~\bibnamefont {Hoek}}, \bibinfo {author} {\bibfnamefont
  {N.}~\bibnamefont {Poccia}}, \bibinfo {author} {\bibfnamefont
  {A.}~\bibnamefont {Vailionis}}, \bibinfo {author} {\bibfnamefont
  {T.}~\bibnamefont {Venkatesan}}, \bibinfo {author} {\bibfnamefont {J.~M.~D.}\
  \bibnamefont {Coey}}, \bibinfo {author} {\bibfnamefont {E.~Y.}\ \bibnamefont
  {Tsymbal}}, \bibinfo {author} {\bibnamefont {Ariando}}, \ and\ \bibinfo
  {author} {\bibfnamefont {H.}~\bibnamefont {Hilgenkamp}},\ }\href {\doibase
  10.1126/science.aaa5198} {\bibfield  {journal} {\bibinfo  {journal}
  {Science}\ }\textbf {\bibinfo {volume} {349}},\ \bibinfo {pages} {716}
  (\bibinfo {year} {2015})}\BibitemShut {NoStop}%
\bibitem [{\citenamefont {Kim}\ and\ \citenamefont
  {Christen}(2010)}]{Kim_2010}%
  \BibitemOpen
  \bibfield  {author} {\bibinfo {author} {\bibfnamefont {H.~S.}\ \bibnamefont
  {Kim}}\ and\ \bibinfo {author} {\bibfnamefont {H.~M.}\ \bibnamefont
  {Christen}},\ }\href {\doibase 10.1088/0953-8984/22/14/146007} {\bibfield
  {journal} {\bibinfo  {journal} {J. Phys.: Condens. Matter.}\ }\textbf
  {\bibinfo {volume} {22}},\ \bibinfo {pages} {146007} (\bibinfo {year}
  {2010})}\BibitemShut {NoStop}%
\bibitem [{\citenamefont {Roqueta}\ \emph {et~al.}(2015)\citenamefont
  {Roqueta}, \citenamefont {Pomar}, \citenamefont {Balcells}, \citenamefont
  {Frontera}, \citenamefont {Valencia}, \citenamefont {Abrudan}, \citenamefont
  {Bozzo}, \citenamefont {Konstantinović}, \citenamefont {Santiso},\ and\
  \citenamefont {Martínez}}]{roqueta2015strain}%
  \BibitemOpen
  \bibfield  {author} {\bibinfo {author} {\bibfnamefont {J.}~\bibnamefont
  {Roqueta}}, \bibinfo {author} {\bibfnamefont {A.}~\bibnamefont {Pomar}},
  \bibinfo {author} {\bibfnamefont {L.}~\bibnamefont {Balcells}}, \bibinfo
  {author} {\bibfnamefont {C.}~\bibnamefont {Frontera}}, \bibinfo {author}
  {\bibfnamefont {S.}~\bibnamefont {Valencia}}, \bibinfo {author}
  {\bibfnamefont {R.}~\bibnamefont {Abrudan}}, \bibinfo {author} {\bibfnamefont
  {B.}~\bibnamefont {Bozzo}}, \bibinfo {author} {\bibfnamefont
  {Z.}~\bibnamefont {Konstantinović}}, \bibinfo {author} {\bibfnamefont
  {J.}~\bibnamefont {Santiso}}, \ and\ \bibinfo {author} {\bibfnamefont
  {B.}~\bibnamefont {Martínez}},\ }\href {\doibase 10.1021/acs.cgd.5b00884}
  {\bibfield  {journal} {\bibinfo  {journal} {Cryst. Growth Des.}\ }\textbf
  {\bibinfo {volume} {15}},\ \bibinfo {pages} {5332} (\bibinfo {year}
  {2015})}\BibitemShut {NoStop}%
\bibitem [{\citenamefont {Choi}\ \emph {et~al.}(2009)\citenamefont {Choi},
  \citenamefont {Marton}, \citenamefont {Jang}, \citenamefont {Moon},
  \citenamefont {Jeon}, \citenamefont {Shin}, \citenamefont {Seo},
  \citenamefont {Noh}, \citenamefont {Myung-Whun}, \citenamefont {Lee},\ and\
  \citenamefont {Lee}}]{Choi_2009}%
  \BibitemOpen
  \bibfield  {author} {\bibinfo {author} {\bibfnamefont {W.~S.}\ \bibnamefont
  {Choi}}, \bibinfo {author} {\bibfnamefont {Z.}~\bibnamefont {Marton}},
  \bibinfo {author} {\bibfnamefont {S.~Y.}\ \bibnamefont {Jang}}, \bibinfo
  {author} {\bibfnamefont {S.~J.}\ \bibnamefont {Moon}}, \bibinfo {author}
  {\bibfnamefont {B.~C.}\ \bibnamefont {Jeon}}, \bibinfo {author}
  {\bibfnamefont {J.~H.}\ \bibnamefont {Shin}}, \bibinfo {author}
  {\bibfnamefont {S.~S.~A.}\ \bibnamefont {Seo}}, \bibinfo {author}
  {\bibfnamefont {T.~W.}\ \bibnamefont {Noh}}, \bibinfo {author} {\bibfnamefont
  {K.}~\bibnamefont {Myung-Whun}}, \bibinfo {author} {\bibfnamefont {H.~N.}\
  \bibnamefont {Lee}}, \ and\ \bibinfo {author} {\bibfnamefont {Y.~S.}\
  \bibnamefont {Lee}},\ }\href {\doibase 10.1088/0022-3727/42/16/165401}
  {\bibfield  {journal} {\bibinfo  {journal} {J. Phys. D Appl. Phys.}\ }\textbf
  {\bibinfo {volume} {42}},\ \bibinfo {pages} {165401} (\bibinfo {year}
  {2009})}\BibitemShut {NoStop}%
\bibitem [{\citenamefont {Adler}(2001)}]{Adler2001}%
  \BibitemOpen
  \bibfield  {author} {\bibinfo {author} {\bibfnamefont {S.~B.}\ \bibnamefont
  {Adler}},\ }\href {\doibase 10.1111/j.1151-2916.2001.tb00968.x} {\bibfield
  {journal} {\bibinfo  {journal} {J. Am. Ceram. Soc.}\ }\textbf {\bibinfo
  {volume} {84}},\ \bibinfo {pages} {2117} (\bibinfo {year}
  {2001})}\BibitemShut {NoStop}%
\bibitem [{\citenamefont {Aschauer}\ \emph {et~al.}(2013)\citenamefont
  {Aschauer}, \citenamefont {Pfenninger}, \citenamefont {Selbach},
  \citenamefont {Grande},\ and\ \citenamefont {Spaldin}}]{Aschauer2013}%
  \BibitemOpen
  \bibfield  {author} {\bibinfo {author} {\bibfnamefont {U.}~\bibnamefont
  {Aschauer}}, \bibinfo {author} {\bibfnamefont {R.}~\bibnamefont
  {Pfenninger}}, \bibinfo {author} {\bibfnamefont {S.~M.}\ \bibnamefont
  {Selbach}}, \bibinfo {author} {\bibfnamefont {T.}~\bibnamefont {Grande}}, \
  and\ \bibinfo {author} {\bibfnamefont {N.~A.}\ \bibnamefont {Spaldin}},\
  }\href {\doibase 10.1103/PhysRevB.88.054111} {\bibfield  {journal} {\bibinfo
  {journal} {Phys. Rev. B}\ }\textbf {\bibinfo {volume} {88}},\ \bibinfo
  {pages} {054111} (\bibinfo {year} {2013})}\BibitemShut {NoStop}%
\bibitem [{\citenamefont {Marthinsen}\ \emph {et~al.}(2016)\citenamefont
  {Marthinsen}, \citenamefont {Faber}, \citenamefont {Aschauer}, \citenamefont
  {Spaldin},\ and\ \citenamefont {Selbach}}]{marthinsen2016}%
  \BibitemOpen
  \bibfield  {author} {\bibinfo {author} {\bibfnamefont {A.}~\bibnamefont
  {Marthinsen}}, \bibinfo {author} {\bibfnamefont {C.}~\bibnamefont {Faber}},
  \bibinfo {author} {\bibfnamefont {U.}~\bibnamefont {Aschauer}}, \bibinfo
  {author} {\bibfnamefont {N.~A.}\ \bibnamefont {Spaldin}}, \ and\ \bibinfo
  {author} {\bibfnamefont {S.~M.}\ \bibnamefont {Selbach}},\ }\href {\doibase
  10.1557/mrc.2016.30} {\bibfield  {journal} {\bibinfo  {journal} {MRS
  Commun.}\ }\textbf {\bibinfo {volume} {6}},\ \bibinfo {pages} {182} (\bibinfo
  {year} {2016})}\BibitemShut {NoStop}%
\bibitem [{\citenamefont {Ricca}\ \emph {et~al.}(2019)\citenamefont {Ricca},
  \citenamefont {Timrov}, \citenamefont {Cococcioni}, \citenamefont {Marzari},\
  and\ \citenamefont {Aschauer}}]{Ricca2019}%
  \BibitemOpen
  \bibfield  {author} {\bibinfo {author} {\bibfnamefont {C.}~\bibnamefont
  {Ricca}}, \bibinfo {author} {\bibfnamefont {I.}~\bibnamefont {Timrov}},
  \bibinfo {author} {\bibfnamefont {M.}~\bibnamefont {Cococcioni}}, \bibinfo
  {author} {\bibfnamefont {N.}~\bibnamefont {Marzari}}, \ and\ \bibinfo
  {author} {\bibfnamefont {U.}~\bibnamefont {Aschauer}},\ }\href {\doibase
  10.1103/PhysRevB.99.094102} {\bibfield  {journal} {\bibinfo  {journal} {Phys.
  Rev. B}\ }\textbf {\bibinfo {volume} {99}},\ \bibinfo {pages} {094102}
  (\bibinfo {year} {2019})}\BibitemShut {NoStop}%
\bibitem [{\citenamefont {Elemans}\ \emph {et~al.}(1971)\citenamefont
  {Elemans}, \citenamefont {Van~Laar}, \citenamefont {Van Der~Veen},\ and\
  \citenamefont {Loopstra}}]{ELEMANS1971238}%
  \BibitemOpen
  \bibfield  {author} {\bibinfo {author} {\bibfnamefont {J.~B. A.~A.}\
  \bibnamefont {Elemans}}, \bibinfo {author} {\bibfnamefont {B.}~\bibnamefont
  {Van~Laar}}, \bibinfo {author} {\bibfnamefont {K.~R.}\ \bibnamefont {Van
  Der~Veen}}, \ and\ \bibinfo {author} {\bibfnamefont {B.~O.}\ \bibnamefont
  {Loopstra}},\ }\href {\doibase https://doi.org/10.1016/0022-4596(71)90034-X}
  {\bibfield  {journal} {\bibinfo  {journal} {J. Solid State Chem.}\ }\textbf
  {\bibinfo {volume} {3}},\ \bibinfo {pages} {238} (\bibinfo {year}
  {1971})}\BibitemShut {NoStop}%
\bibitem [{\citenamefont {Arima}\ \emph {et~al.}(1993)\citenamefont {Arima},
  \citenamefont {Tokura},\ and\ \citenamefont {Torrance}}]{Arima1993}%
  \BibitemOpen
  \bibfield  {author} {\bibinfo {author} {\bibfnamefont {T.}~\bibnamefont
  {Arima}}, \bibinfo {author} {\bibfnamefont {Y.}~\bibnamefont {Tokura}}, \
  and\ \bibinfo {author} {\bibfnamefont {J.~B.}\ \bibnamefont {Torrance}},\
  }\href {\doibase 10.1103/PhysRevB.48.17006} {\bibfield  {journal} {\bibinfo
  {journal} {Phys. Rev. B}\ }\textbf {\bibinfo {volume} {48}},\ \bibinfo
  {pages} {17006} (\bibinfo {year} {1993})}\BibitemShut {NoStop}%
\bibitem [{\citenamefont {Mahendiran}\ \emph {et~al.}(1995)\citenamefont
  {Mahendiran}, \citenamefont {Raychaudhuri}, \citenamefont {Chainani},
  \citenamefont {Sarma},\ and\ \citenamefont {Roy}}]{Mahendiran1995}%
  \BibitemOpen
  \bibfield  {author} {\bibinfo {author} {\bibfnamefont {R.}~\bibnamefont
  {Mahendiran}}, \bibinfo {author} {\bibfnamefont {A.~K.}\ \bibnamefont
  {Raychaudhuri}}, \bibinfo {author} {\bibfnamefont {A.}~\bibnamefont
  {Chainani}}, \bibinfo {author} {\bibfnamefont {D.~D.}\ \bibnamefont {Sarma}},
  \ and\ \bibinfo {author} {\bibfnamefont {S.~B.}\ \bibnamefont {Roy}},\ }\href
  {\doibase 10.1063/1.113556} {\bibfield  {journal} {\bibinfo  {journal} {Appl.
  Phys. Lett.}\ }\textbf {\bibinfo {volume} {66}},\ \bibinfo {pages} {233}
  (\bibinfo {year} {1995})}\BibitemShut {NoStop}%
\bibitem [{\citenamefont {Giannozzi}\ \emph {et~al.}(2009)\citenamefont
  {Giannozzi}, \citenamefont {Baroni}, \citenamefont {Bonini}, \citenamefont
  {Calandra}, \citenamefont {Car}, \citenamefont {Cavazzoni}, \citenamefont
  {Ceresoli}, \citenamefont {Chiarotti}, \citenamefont {Cococcioni},
  \citenamefont {Dabo}, \citenamefont {Corso}, \citenamefont {de~Gironcoli},
  \citenamefont {Fabris}, \citenamefont {Fratesi}, \citenamefont {Gebauer},
  \citenamefont {Gerstmann}, \citenamefont {Gougoussis}, \citenamefont
  {Kokalj}, \citenamefont {Lazzeri}, \citenamefont {Martin-Samos},
  \citenamefont {Marzari}, \citenamefont {Mauri}, \citenamefont {Mazzarello},
  \citenamefont {Paolini}, \citenamefont {Pasquarello}, \citenamefont
  {Paulatto}, \citenamefont {Sbraccia}, \citenamefont {Scandolo}, \citenamefont
  {Sclauzero}, \citenamefont {Seitsonen}, \citenamefont {Smogunov},
  \citenamefont {Umari},\ and\ \citenamefont
  {Wentzcovitch}}]{giannozzi2009quantum}%
  \BibitemOpen
  \bibfield  {author} {\bibinfo {author} {\bibfnamefont {P.}~\bibnamefont
  {Giannozzi}}, \bibinfo {author} {\bibfnamefont {S.}~\bibnamefont {Baroni}},
  \bibinfo {author} {\bibfnamefont {N.}~\bibnamefont {Bonini}}, \bibinfo
  {author} {\bibfnamefont {M.}~\bibnamefont {Calandra}}, \bibinfo {author}
  {\bibfnamefont {R.}~\bibnamefont {Car}}, \bibinfo {author} {\bibfnamefont
  {C.}~\bibnamefont {Cavazzoni}}, \bibinfo {author} {\bibfnamefont
  {D.}~\bibnamefont {Ceresoli}}, \bibinfo {author} {\bibfnamefont {G.~L.}\
  \bibnamefont {Chiarotti}}, \bibinfo {author} {\bibfnamefont {M.}~\bibnamefont
  {Cococcioni}}, \bibinfo {author} {\bibfnamefont {I.}~\bibnamefont {Dabo}},
  \bibinfo {author} {\bibfnamefont {A.~D.}\ \bibnamefont {Corso}}, \bibinfo
  {author} {\bibfnamefont {S.}~\bibnamefont {de~Gironcoli}}, \bibinfo {author}
  {\bibfnamefont {S.}~\bibnamefont {Fabris}}, \bibinfo {author} {\bibfnamefont
  {G.}~\bibnamefont {Fratesi}}, \bibinfo {author} {\bibfnamefont
  {R.}~\bibnamefont {Gebauer}}, \bibinfo {author} {\bibfnamefont
  {U.}~\bibnamefont {Gerstmann}}, \bibinfo {author} {\bibfnamefont
  {C.}~\bibnamefont {Gougoussis}}, \bibinfo {author} {\bibfnamefont
  {A.}~\bibnamefont {Kokalj}}, \bibinfo {author} {\bibfnamefont
  {M.}~\bibnamefont {Lazzeri}}, \bibinfo {author} {\bibfnamefont
  {L.}~\bibnamefont {Martin-Samos}}, \bibinfo {author} {\bibfnamefont
  {N.}~\bibnamefont {Marzari}}, \bibinfo {author} {\bibfnamefont
  {F.}~\bibnamefont {Mauri}}, \bibinfo {author} {\bibfnamefont
  {R.}~\bibnamefont {Mazzarello}}, \bibinfo {author} {\bibfnamefont
  {S.}~\bibnamefont {Paolini}}, \bibinfo {author} {\bibfnamefont
  {A.}~\bibnamefont {Pasquarello}}, \bibinfo {author} {\bibfnamefont
  {L.}~\bibnamefont {Paulatto}}, \bibinfo {author} {\bibfnamefont
  {C.}~\bibnamefont {Sbraccia}}, \bibinfo {author} {\bibfnamefont
  {S.}~\bibnamefont {Scandolo}}, \bibinfo {author} {\bibfnamefont
  {G.}~\bibnamefont {Sclauzero}}, \bibinfo {author} {\bibfnamefont {A.~P.}\
  \bibnamefont {Seitsonen}}, \bibinfo {author} {\bibfnamefont {A.}~\bibnamefont
  {Smogunov}}, \bibinfo {author} {\bibfnamefont {P.}~\bibnamefont {Umari}}, \
  and\ \bibinfo {author} {\bibfnamefont {R.~M.}\ \bibnamefont {Wentzcovitch}},\
  }\href {\doibase 10.1088/0953-8984/21/39/395502} {\bibfield  {journal}
  {\bibinfo  {journal} {J. Phys.: Condens. Matter.}\ }\textbf {\bibinfo
  {volume} {21}},\ \bibinfo {pages} {395502} (\bibinfo {year}
  {2009})}\BibitemShut {NoStop}%
\bibitem [{\citenamefont {Giannozzi}\ \emph {et~al.}(2017)\citenamefont
  {Giannozzi}, \citenamefont {Andreussi}, \citenamefont {Brumme}, \citenamefont
  {Bunau}, \citenamefont {{Buongiorno Nardelli}}, \citenamefont {Calandra},
  \citenamefont {Car}, \citenamefont {Cavazzoni}, \citenamefont {Ceresoli},
  \citenamefont {Cococcioni}, \citenamefont {Colonna}, \citenamefont
  {Carnimeo}, \citenamefont {{Dal Corso}}, \citenamefont {{De Gironcoli}},
  \citenamefont {Delugas}, \citenamefont {Distasio}, \citenamefont {Ferretti},
  \citenamefont {Floris}, \citenamefont {Fratesi}, \citenamefont {Fugallo},
  \citenamefont {Gebauer}, \citenamefont {Gerstmann}, \citenamefont {Giustino},
  \citenamefont {Gorni}, \citenamefont {Jia}, \citenamefont {Kawamura},
  \citenamefont {Ko}, \citenamefont {Kokalj}, \citenamefont
  {K{\"{u}}c{\"{u}}kbenli}, \citenamefont {Lazzeri}, \citenamefont {Marsili},
  \citenamefont {Marzari}, \citenamefont {Mauri}, \citenamefont {Nguyen},
  \citenamefont {Nguyen}, \citenamefont {Otero-De-La-Roza}, \citenamefont
  {Paulatto}, \citenamefont {Ponc{\'{e}}}, \citenamefont {Rocca}, \citenamefont
  {Sabatini}, \citenamefont {Santra}, \citenamefont {Schlipf}, \citenamefont
  {Seitsonen}, \citenamefont {Smogunov}, \citenamefont {Timrov}, \citenamefont
  {Thonhauser}, \citenamefont {Umari}, \citenamefont {Vast}, \citenamefont
  {Wu},\ and\ \citenamefont {Baroni}}]{Giannozzi2017}%
  \BibitemOpen
  \bibfield  {author} {\bibinfo {author} {\bibfnamefont {P.}~\bibnamefont
  {Giannozzi}}, \bibinfo {author} {\bibfnamefont {O.}~\bibnamefont
  {Andreussi}}, \bibinfo {author} {\bibfnamefont {T.}~\bibnamefont {Brumme}},
  \bibinfo {author} {\bibfnamefont {O.}~\bibnamefont {Bunau}}, \bibinfo
  {author} {\bibfnamefont {M.}~\bibnamefont {{Buongiorno Nardelli}}}, \bibinfo
  {author} {\bibfnamefont {M.}~\bibnamefont {Calandra}}, \bibinfo {author}
  {\bibfnamefont {R.}~\bibnamefont {Car}}, \bibinfo {author} {\bibfnamefont
  {C.}~\bibnamefont {Cavazzoni}}, \bibinfo {author} {\bibfnamefont
  {D.}~\bibnamefont {Ceresoli}}, \bibinfo {author} {\bibfnamefont
  {M.}~\bibnamefont {Cococcioni}}, \bibinfo {author} {\bibfnamefont
  {N.}~\bibnamefont {Colonna}}, \bibinfo {author} {\bibfnamefont
  {I.}~\bibnamefont {Carnimeo}}, \bibinfo {author} {\bibfnamefont
  {A.}~\bibnamefont {{Dal Corso}}}, \bibinfo {author} {\bibfnamefont
  {S.}~\bibnamefont {{De Gironcoli}}}, \bibinfo {author} {\bibfnamefont
  {P.}~\bibnamefont {Delugas}}, \bibinfo {author} {\bibfnamefont {R.~A.}\
  \bibnamefont {Distasio}}, \bibinfo {author} {\bibfnamefont {A.}~\bibnamefont
  {Ferretti}}, \bibinfo {author} {\bibfnamefont {A.}~\bibnamefont {Floris}},
  \bibinfo {author} {\bibfnamefont {G.}~\bibnamefont {Fratesi}}, \bibinfo
  {author} {\bibfnamefont {G.}~\bibnamefont {Fugallo}}, \bibinfo {author}
  {\bibfnamefont {R.}~\bibnamefont {Gebauer}}, \bibinfo {author} {\bibfnamefont
  {U.}~\bibnamefont {Gerstmann}}, \bibinfo {author} {\bibfnamefont
  {F.}~\bibnamefont {Giustino}}, \bibinfo {author} {\bibfnamefont
  {T.}~\bibnamefont {Gorni}}, \bibinfo {author} {\bibfnamefont
  {J.}~\bibnamefont {Jia}}, \bibinfo {author} {\bibfnamefont {M.}~\bibnamefont
  {Kawamura}}, \bibinfo {author} {\bibfnamefont {H.~Y.}\ \bibnamefont {Ko}},
  \bibinfo {author} {\bibfnamefont {A.}~\bibnamefont {Kokalj}}, \bibinfo
  {author} {\bibfnamefont {E.}~\bibnamefont {K{\"{u}}c{\"{u}}kbenli}}, \bibinfo
  {author} {\bibfnamefont {M.}~\bibnamefont {Lazzeri}}, \bibinfo {author}
  {\bibfnamefont {M.}~\bibnamefont {Marsili}}, \bibinfo {author} {\bibfnamefont
  {N.}~\bibnamefont {Marzari}}, \bibinfo {author} {\bibfnamefont
  {F.}~\bibnamefont {Mauri}}, \bibinfo {author} {\bibfnamefont {N.~L.}\
  \bibnamefont {Nguyen}}, \bibinfo {author} {\bibfnamefont {H.~V.}\
  \bibnamefont {Nguyen}}, \bibinfo {author} {\bibfnamefont {A.}~\bibnamefont
  {Otero-De-La-Roza}}, \bibinfo {author} {\bibfnamefont {L.}~\bibnamefont
  {Paulatto}}, \bibinfo {author} {\bibfnamefont {S.}~\bibnamefont
  {Ponc{\'{e}}}}, \bibinfo {author} {\bibfnamefont {D.}~\bibnamefont {Rocca}},
  \bibinfo {author} {\bibfnamefont {R.}~\bibnamefont {Sabatini}}, \bibinfo
  {author} {\bibfnamefont {B.}~\bibnamefont {Santra}}, \bibinfo {author}
  {\bibfnamefont {M.}~\bibnamefont {Schlipf}}, \bibinfo {author} {\bibfnamefont
  {A.~P.}\ \bibnamefont {Seitsonen}}, \bibinfo {author} {\bibfnamefont
  {A.}~\bibnamefont {Smogunov}}, \bibinfo {author} {\bibfnamefont
  {I.}~\bibnamefont {Timrov}}, \bibinfo {author} {\bibfnamefont
  {T.}~\bibnamefont {Thonhauser}}, \bibinfo {author} {\bibfnamefont
  {P.}~\bibnamefont {Umari}}, \bibinfo {author} {\bibfnamefont
  {N.}~\bibnamefont {Vast}}, \bibinfo {author} {\bibfnamefont {X.}~\bibnamefont
  {Wu}}, \ and\ \bibinfo {author} {\bibfnamefont {S.}~\bibnamefont {Baroni}},\
  }\href {\doibase 10.1088/1361-648X/aa8f79} {\bibfield  {journal} {\bibinfo
  {journal} {J. Phys.: Condens. Matter.}\ }\textbf {\bibinfo {volume} {29}},\
  \bibinfo {pages} {465901} (\bibinfo {year} {2017})}\BibitemShut {NoStop}%
\bibitem [{\citenamefont {Perdew}\ \emph {et~al.}(2008)\citenamefont {Perdew},
  \citenamefont {Ruzsinszky}, \citenamefont {Csonka}, \citenamefont {Vydrov},
  \citenamefont {Scuseria}, \citenamefont {Constantin}, \citenamefont {Zhou},\
  and\ \citenamefont {Burke}}]{perdew2008pbesol}%
  \BibitemOpen
  \bibfield  {author} {\bibinfo {author} {\bibfnamefont {J.~P.}\ \bibnamefont
  {Perdew}}, \bibinfo {author} {\bibfnamefont {A.}~\bibnamefont {Ruzsinszky}},
  \bibinfo {author} {\bibfnamefont {G.~I.}\ \bibnamefont {Csonka}}, \bibinfo
  {author} {\bibfnamefont {O.~A.}\ \bibnamefont {Vydrov}}, \bibinfo {author}
  {\bibfnamefont {G.~E.}\ \bibnamefont {Scuseria}}, \bibinfo {author}
  {\bibfnamefont {L.~A.}\ \bibnamefont {Constantin}}, \bibinfo {author}
  {\bibfnamefont {X.}~\bibnamefont {Zhou}}, \ and\ \bibinfo {author}
  {\bibfnamefont {K.}~\bibnamefont {Burke}},\ }\href {\doibase
  10.1103/PhysRevLett.100.136406} {\bibfield  {journal} {\bibinfo  {journal}
  {Phys. Rev. Lett.}\ }\textbf {\bibinfo {volume} {100}},\ \bibinfo {pages}
  {136406} (\bibinfo {year} {2008})}\BibitemShut {NoStop}%
\bibitem [{\citenamefont {Dudarev}\ \emph {et~al.}(1998)\citenamefont
  {Dudarev}, \citenamefont {Botton}, \citenamefont {Savrasov}, \citenamefont
  {Humphreys},\ and\ \citenamefont {Sutton}}]{Dudarev1998}%
  \BibitemOpen
  \bibfield  {author} {\bibinfo {author} {\bibfnamefont {S.~L.}\ \bibnamefont
  {Dudarev}}, \bibinfo {author} {\bibfnamefont {G.~A.}\ \bibnamefont {Botton}},
  \bibinfo {author} {\bibfnamefont {S.~Y.}\ \bibnamefont {Savrasov}}, \bibinfo
  {author} {\bibfnamefont {C.~J.}\ \bibnamefont {Humphreys}}, \ and\ \bibinfo
  {author} {\bibfnamefont {A.~P.}\ \bibnamefont {Sutton}},\ }\href {\doibase
  10.1103/PhysRevB.57.1505} {\bibfield  {journal} {\bibinfo  {journal} {Phys.
  Rev. B}\ }\textbf {\bibinfo {volume} {57}},\ \bibinfo {pages} {1505}
  (\bibinfo {year} {1998})}\BibitemShut {NoStop}%
\bibitem [{\citenamefont {Timrov}\ \emph {et~al.}(2018)\citenamefont {Timrov},
  \citenamefont {Marzari},\ and\ \citenamefont {Cococcioni}}]{Timrov2018}%
  \BibitemOpen
  \bibfield  {author} {\bibinfo {author} {\bibfnamefont {I.}~\bibnamefont
  {Timrov}}, \bibinfo {author} {\bibfnamefont {N.}~\bibnamefont {Marzari}}, \
  and\ \bibinfo {author} {\bibfnamefont {M.}~\bibnamefont {Cococcioni}},\
  }\href {\doibase 10.1103/PhysRevB.98.085127} {\bibfield  {journal} {\bibinfo
  {journal} {Phys. Rev. B}\ }\textbf {\bibinfo {volume} {98}},\ \bibinfo
  {pages} {085127} (\bibinfo {year} {2018})}\BibitemShut {NoStop}%
\bibitem [{\citenamefont {Hashimoto}\ \emph {et~al.}(2010)\citenamefont
  {Hashimoto}, \citenamefont {Ishibashi},\ and\ \citenamefont
  {Terakura}}]{Hashimoto2010}%
  \BibitemOpen
  \bibfield  {author} {\bibinfo {author} {\bibfnamefont {T.}~\bibnamefont
  {Hashimoto}}, \bibinfo {author} {\bibfnamefont {S.}~\bibnamefont
  {Ishibashi}}, \ and\ \bibinfo {author} {\bibfnamefont {K.}~\bibnamefont
  {Terakura}},\ }\href {\doibase 10.1103/PhysRevB.82.045124} {\bibfield
  {journal} {\bibinfo  {journal} {Phys. Rev. B}\ }\textbf {\bibinfo {volume}
  {82}},\ \bibinfo {pages} {045124} (\bibinfo {year} {2010})}\BibitemShut
  {NoStop}%
\bibitem [{\citenamefont {He}\ and\ \citenamefont
  {Franchini}(2012)}]{Franchini201286}%
  \BibitemOpen
  \bibfield  {author} {\bibinfo {author} {\bibfnamefont {J.}~\bibnamefont
  {He}}\ and\ \bibinfo {author} {\bibfnamefont {C.}~\bibnamefont {Franchini}},\
  }\href {\doibase 10.1103/PhysRevB.86.235117} {\bibfield  {journal} {\bibinfo
  {journal} {Phys. Rev. B}\ }\textbf {\bibinfo {volume} {86}},\ \bibinfo
  {pages} {235117} (\bibinfo {year} {2012})}\BibitemShut {NoStop}%
\bibitem [{\citenamefont {Gavin}\ and\ \citenamefont
  {Watson}(2017)}]{GAVIN201713}%
  \BibitemOpen
  \bibfield  {author} {\bibinfo {author} {\bibfnamefont {A.~L.}\ \bibnamefont
  {Gavin}}\ and\ \bibinfo {author} {\bibfnamefont {G.~W.}\ \bibnamefont
  {Watson}},\ }\href {\doibase https://doi.org/10.1016/j.ssi.2016.10.007}
  {\bibfield  {journal} {\bibinfo  {journal} {Solid State Ion.}\ }\textbf
  {\bibinfo {volume} {299}},\ \bibinfo {pages} {13} (\bibinfo {year}
  {2017})}\BibitemShut {NoStop}%
\bibitem [{\citenamefont {Kr{\"o}ger}\ and\ \citenamefont
  {Vink}(1956)}]{KROGER1956307}%
  \BibitemOpen
  \bibfield  {author} {\bibinfo {author} {\bibfnamefont {F.~A.}\ \bibnamefont
  {Kr{\"o}ger}}\ and\ \bibinfo {author} {\bibfnamefont {H.~J.}\ \bibnamefont
  {Vink}},\ }\href {\doibase 10.1016/S0081-1947(08)60135-6} {\bibfield
  {journal} {\bibinfo  {journal} {Solid State Phys.}\ }\textbf {\bibinfo
  {volume} {3}},\ \bibinfo {pages} {307} (\bibinfo {year} {1956})}\BibitemShut
  {NoStop}%
\bibitem [{\citenamefont {Pavone}\ \emph {et~al.}(2014)\citenamefont {Pavone},
  \citenamefont {Muñoz-García}, \citenamefont {Ritzmann},\ and\ \citenamefont
  {Carter}}]{Pavone2014}%
  \BibitemOpen
  \bibfield  {author} {\bibinfo {author} {\bibfnamefont {M.}~\bibnamefont
  {Pavone}}, \bibinfo {author} {\bibfnamefont {A.~B.}\ \bibnamefont
  {Muñoz-García}}, \bibinfo {author} {\bibfnamefont {A.~M.}\ \bibnamefont
  {Ritzmann}}, \ and\ \bibinfo {author} {\bibfnamefont {E.~A.}\ \bibnamefont
  {Carter}},\ }\href {\doibase 10.1021/jp500352h} {\bibfield  {journal}
  {\bibinfo  {journal} {J. Phys. Chem. C}\ }\textbf {\bibinfo {volume} {118}},\
  \bibinfo {pages} {13346} (\bibinfo {year} {2014})}\BibitemShut {NoStop}%
\bibitem [{\citenamefont {Olsson}\ \emph {et~al.}(2016)\citenamefont {Olsson},
  \citenamefont {Aparicio-Anglès},\ and\ \citenamefont
  {de~Leeuw}}]{Olsson2016}%
  \BibitemOpen
  \bibfield  {author} {\bibinfo {author} {\bibfnamefont {E.}~\bibnamefont
  {Olsson}}, \bibinfo {author} {\bibfnamefont {X.}~\bibnamefont
  {Aparicio-Anglès}}, \ and\ \bibinfo {author} {\bibfnamefont {N.~H.}\
  \bibnamefont {de~Leeuw}},\ }\href {\doibase 10.1063/1.4954939} {\bibfield
  {journal} {\bibinfo  {journal} {J. Chem. Phys.}\ }\textbf {\bibinfo {volume}
  {145}},\ \bibinfo {pages} {014703} (\bibinfo {year} {2016})}\BibitemShut
  {NoStop}%
\bibitem [{\citenamefont {Sit}\ \emph {et~al.}(2011)\citenamefont {Sit},
  \citenamefont {Car}, \citenamefont {Cohen},\ and\ \citenamefont
  {Selloni}}]{Sit2011}%
  \BibitemOpen
  \bibfield  {author} {\bibinfo {author} {\bibfnamefont {P.~H.-L.}\
  \bibnamefont {Sit}}, \bibinfo {author} {\bibfnamefont {R.}~\bibnamefont
  {Car}}, \bibinfo {author} {\bibfnamefont {M.~H.}\ \bibnamefont {Cohen}}, \
  and\ \bibinfo {author} {\bibfnamefont {A.}~\bibnamefont {Selloni}},\ }\href
  {\doibase 10.1021/ic2013107} {\bibfield  {journal} {\bibinfo  {journal}
  {Inorg. Chem.}\ }\textbf {\bibinfo {volume} {50}},\ \bibinfo {pages} {10259}
  (\bibinfo {year} {2011})}\BibitemShut {NoStop}%
\bibitem [{\citenamefont {Aschauer}\ and\ \citenamefont
  {Spaldin}(2016)}]{aschauer2016interplay}%
  \BibitemOpen
  \bibfield  {author} {\bibinfo {author} {\bibfnamefont {U.}~\bibnamefont
  {Aschauer}}\ and\ \bibinfo {author} {\bibfnamefont {N.~A.}\ \bibnamefont
  {Spaldin}},\ }\href {\doibase 10.1063/1.4958716} {\bibfield  {journal}
  {\bibinfo  {journal} {Appl. Phys. Lett.}\ }\textbf {\bibinfo {volume}
  {109}},\ \bibinfo {pages} {031901} (\bibinfo {year} {2016})}\BibitemShut
  {NoStop}%
\end{thebibliography}%


\clearpage
\clearpage 
\setcounter{page}{1}
\renewcommand{\thetable}{S\arabic{table}}  
\setcounter{table}{0}
\renewcommand{\thefigure}{S\arabic{figure}}
\setcounter{figure}{0}
\renewcommand{\thesection}{S\arabic{section}}
\setcounter{section}{0}
\renewcommand{\theequation}{S\arabic{equation}}
\setcounter{equation}{0}
\onecolumngrid

\begin{center}
\textbf{Supplementary information for\\\vspace{0.5 cm}
\large Local polarization in oxygen-deficient LaMnO$_3$ induced by charge localization in the Jahn-Teller distorted structure\\\vspace{0.3 cm}}
Chiara Ricca$^{1, 2}$, Nicolas Niederhauser,$^{1,2}$
and Ulrich Aschauer$^{1, 2}$

\small
$^1$\textit{Department of Chemistry and Biochemistry, University of Bern, Freiestrasse 3, CH-3012 Bern, Switzerland}

$^2$\textit{National Centre for Computational Design and Discovery of Novel Materials (MARVEL), Switzerland}

(Dated: \today)
\end{center}


\section{Methods}
\label{sec:compdetails}
The calculations were performed in the framework of DFT with the {\sc{Quantum ESPRESSO}} package~\citeSI{SI_giannozzi2009quantum,SI_Giannozzi2017}. PBEsol~\citeSI{SI_perdew2008pbesol} was used as exchange-correlation functional together with ultrasoft pseudopotentials~ \citeSI{SI_vanderbilt1990soft} including La($5s$,$5p$,$5d$,$6s$,$6p$), Mn($3p$,$4s$,$3d$), and O($2s$,$2p$) states. A kinetic-energy cutoff of 90 Ry for wave functions and of 1080 Ry for spin-charge density and potentials were applied. A gaussian smearing with a broadening of 0.01 Ry was used in all calculations.

LMO (space group $Pbnm$) was studied using a 40-atom supercell of the 5-atom primitive cubic cell. A  4$\times$4$\times$4 Monkhorst-Pack \textbf{k}-point grid was applied to sample the Brillouin zone of this cell. A denser 8$\times$8$\times$8 grid was used for plotting the density of states (DOS). Only the A-type antiferromagnetic (AFM) phase with ferromagnetic coupling in the \textit{ac} plane was considered. Neutral oxygen vacancies (V$_\textrm{O}^{\bullet \bullet}$) were created by  removing one O  atom from the supercell. We studied two inequivalent oxygen-vacancy positions: an out-of-plane (OP) configuration  and an in-plane one (IP) in which the oxygen removal breaks Mn--O--Mn bonds along the \textit{b}-axis and  in the \textit{ac}-plane, respectively.
For stoichiometric  bulk and isostatic calculations, both lattice parameters and atomic positions were relaxed. Isostatic strain was applied by expanding or shrinking the cell vectors by the same amount in all directions. Thin film geometries with epitaxial biaxial strain in the \textit{ac}-plane to simulate growth on a cubic substrate were instead computed following the procedure reported in Ref.~\citeSI{SI_Rondinelli:2011jk}. 
Finally, for defective systems, atomic positions were optimized while keeping the lattice vectors fixed at optimized values of the stoichiometric system. In all cases, atomic forces were converged to within 5 $\times$ 10$^{-2}$ eV/\AA, while energies were converged to within 1.4 $ \times$10$^{-5}$ eV.  

The rotationally invariant formulation by Dudarev {\it et al.}~\citeSI{SI_Dudarev1998} was used in all the DFT+$U$ calculations.  We imposed a global Hubbard $U$ parameter ($U_\mathrm{SC}$) on all the Mn-$3d$ states. The $U_\mathrm{SC}$ value of 4.08 eV was  computed via DFPT calculations~\citeSI{SI_Timrov2018} and through the self-consistent procedure introduced in Ref.~\citeSI{SI_Ricca2019} for stoichiometric bulk geometries  of the A-AFM phase.  $\Gamma$ point sampling of the \textbf{q}-space was performed for all DFPT calculations. A convergence threshold of 0.01 eV was applied for the self-consistency of  the $U$ values. Atomic orbitals were used to construct occupation matrices and projectors in the DFT+$U$ scheme.

The defect formation energy for a oxygen vacancy  in its neutral state ($\textrm{E}_{\textrm{f},\textrm{V}_\textrm{O}}$) was computed with the following equation (see Ref.~\citeSI{SI_freysoldt2014first}):
\begin{equation}
\textrm{E}_{\textrm{f},\textrm{V}_\textrm{O}}(\epsilon, \mu_O)=\textrm{E}_{\textrm{tot},\textrm{V}_\textrm{O}}(\epsilon) -\textrm{E}_{\textrm{tot,stoic}}(\epsilon) +\mu_\textrm{O}\,,
\label{eq:formenerg}
\end{equation}
where $\textrm{E}_{\textrm{tot},\textrm{V}_\textrm{O}}$ and $\textrm{E}_{\textrm{tot,stoic}}$ are the total energies of the defective and stoichiometric systems, $\epsilon$ is the applied strain  and $\mu_\textrm{O}$ is the oxygen chemical potential.
We will show results in the oxygen-rich limit, \textit{i.e.} with $\mu_{\textrm{O}}=\frac{1}{2}E_{\mathrm{O_2}}$, $E_{\mathrm{O_2}}$ being the energy of an oxygen molecule.

To reduce the computational cost, the polarization $\vec{P}$ was computed using a point-charge model:
\begin{equation}
\vec{P}=\sum_i \vec{r}_i q_i\,,
\label{eq:polarization1}
\end{equation}
where $\vec{r}_i$ is the position of atom $i$ and $q_i$ is its nominal charge: +3 for La, -2 for O, and +3 or +2 for stoichiometric-like or reduced Mn sites. The charge applied on each Mn was defined on the base of its oxidation state computed through the method introduced by Sit \textit{et al.}~\citeSI{SI_Sit2011}. The polarization being a multivalued quantity, it has been corrected by an integer number of polarization quanta $\vec{Q}$, computed as:
\begin{equation}
\vec{Q}=\frac{e}{V} \begin{bmatrix} a\\b\\c \end{bmatrix}\,,
\label{eq:polarization2}
\end{equation}
with $a$, $b$, and $c$ being the lattice parameters, $V$ the volume of the unit cell, and $e$ the elementary charge. 

\newpage
\FloatBarrier

\section{Electronic Properties}

\begin{figure}[h]
 \centering
 \includegraphics[width=85mm]{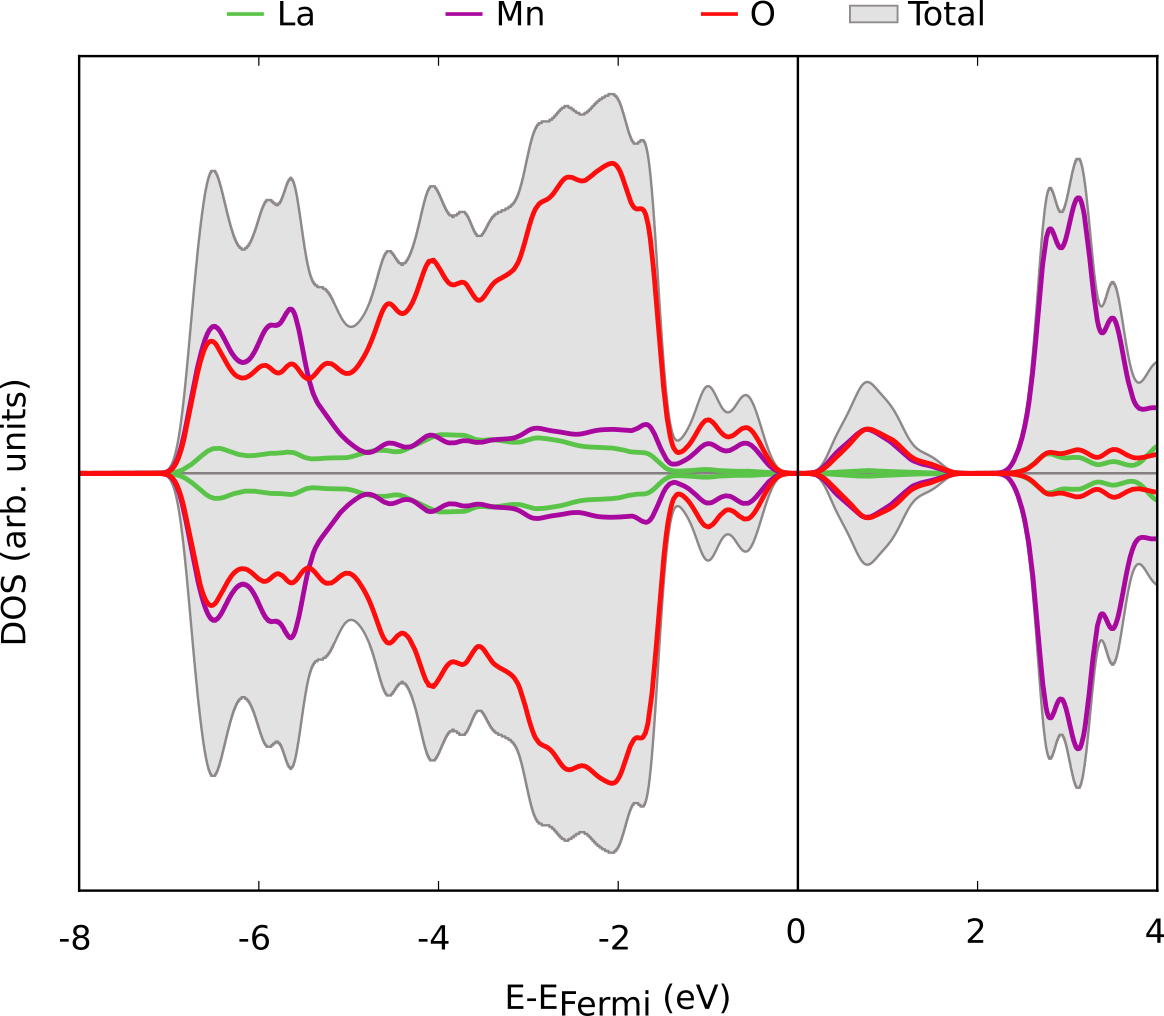}
 \caption{Total and projected density of states (DOS) for bulk LMO in the A-AFM phase. The
zero of the energy scale was set at the Fermi energy value}
\label{fig:PDOS_stoich_unstrained}
\end{figure}

Fig. \ref{fig:PDOS_stoich_unstrained} shows the computed electronic density of states, showing the insulating character of JT-distorted stoichiometric LMO, the Fermi energy lying between the occupied and unoccupied $e_g$ states.

\FloatBarrier
\bibliographystyleSI{apsrev4-1}
\bibliographySI{references}

\end{document}